\documentclass[aip,jmp,amsmath,amssymb,preprint]{revtex4-1}

\usepackage{graphicx}
\usepackage{dcolumn}
\usepackage{bm}
\usepackage{hyperref}

\newcommand{\contr}{\put(0,0){\line(1,0){3}} \put(3,0){\line(0,1){6}}\;}

\begin{document}


\title[]{Lagrangian approach to the 
physical degree of freedom count}


\author{Bogar D\'{\i}az}
\email{bdiaz@alumnos.fcfm.buap.mx}
\affiliation{Facultad de Ciencias F\'{\i}sico Matem\'aticas, Benem\'erita Universidad Aut\'onoma de Puebla, Av. San 
Claudio y 18 Sur, 72570, Puebla, Puebla, M\'exico.}

\author{Daniel Higuita}
\email{dhiguita@fis.cinvestav.mx}
\affiliation{Departamento de F\'{\i}sica, Cinvestav, Instituto Polit\'ecnico Nacional 2508, San Pedro Zacatenco, Gustavo A. 
Madero, Ciudad de M\'exico, M\'exico.}

\author{Merced Montesinos}
\email{merced@fis.cinvestav.mx}
\homepage{http://www.fis.cinvestav.mx/~merced}
\affiliation{Departamento de F\'{\i}sica, Cinvestav, Instituto Polit\'ecnico Nacional 2508, San Pedro Zacatenco, Gustavo A. 
Madero, Ciudad de M\'exico, M\'exico.}

\date{\today}

\begin{abstract}
In this paper we present a Lagrangian method that allows the physical degree of freedom count for any Lagrangian system without having to perform neither Dirac nor covariant canonical analyses. The essence of our method is to establish a map between the relevant Lagrangian parameters of the current approach and the Hamiltonian parameters that enter in the formula for the counting of the physical degrees of freedom as is given in Dirac's method. Once the map is obtained, the usual Hamiltonian formula for the counting can be expressed in terms of Lagrangian parameters only and therefore we can remain in the Lagrangian side without having to go to the Hamiltonian one. Using the map it is also possible to count the number of first and second-class constraints within the Lagrangian formalism only. For the sake of completeness, the geometric structure underlying the current approach--developed for systems with a finite number of degrees of freedom--is uncovered with the help of the covariant canonical formalism. Finally, the method is illustrated in several examples, including the relativistic free particle.
\end{abstract}

\maketitle

\section{\label{intro}Introduction}
To analyze singular Lagrangian systems from the Hamiltonian viewpoint there exists a powerful and well-established approach  known as Dirac's method \cite{Dirac1,Dirac2} (see also Refs.~\onlinecite{Sudarshan, Teitelboim}). This method spoils some desirable features of the theory under study, such as explicit general covariance, for instance. A way to avoid this fact is to perform the analysis of the system in the framework of the covariant canonical formalism.\cite{Wald} As is clear, both approaches have advantages and disadvantages and they complement to each other.

Nevertheless, as practitioners of both approaches we feel that sometimes we need ``something in between'' these two approaches, something that be practical enough (as Dirac's method) but Lagrangian. Something that leads to the right counting of the physical degrees of freedom in the Lagrangian framework but without having to handle all the geometry involved in the covariant canonical formalism. As far as we know, there is not any approach of this kind reported in the literature.

In this paper we report an approach of this kind and, in this sense, we think that we fill out a gap present in the literature of Hamiltonian and Lagrangian systems.

As is well-known, gauge systems (in Dirac's terminology) are particular cases of singular Lagrangian systems. Any Lagrangian formula for the counting of the physical local degrees of freedom must deal with this kind of systems. The formula we report below does it. Our work is just a mixing of ideas previously reported in the literature on the subject, but the formula for the counting is new. It had not been reported before.

(i) The first idea involved in our approach is just the handling of the gauge symmetries in the Lagrangian framework, an issue that goes back to Noether's theorem obviously \cite{Noether} (see also Refs.~\onlinecite{Sudarshan, Shirzad, Rothe} for the purposes of the present paper). Because of this, Section \ref{sec2} contains a summary of Noether's theorem for uncovering the gauge symmetries of the Lagrangian equations of motion with the goal of fixing the notation employed in the rest of the paper too. The details of this construction are contained in the Appendix \ref{appA}. Applications of this procedure can already be found in Refs.~\onlinecite{Rothe1997,Kim1998}.

(ii) The second idea involved in the analysis consists in choosing the relevant Lagrangian parameters.

With ingredients (i) and (ii) at hand,  a Lagrangian formula to count the number of the physical local degrees of freedom is cooked, and it is reported in Section \ref{sec3}, which includes the main result of this paper. It is worth mentioning that an attempt for the formula was reported in Ref.~\onlinecite{Zanelli} for systems having irreducible first-class constraints only. This is quite restrictive and excludes systems with all kind of constraints from the Hamiltonian viewpoint (reducible constraints, second-class constraints, etc.)

The idea to get the formula reported in Section \ref{sec3} is very simple: it lies in relating the relevant parameters of the Lagrangian and Hamiltonian approaches employing two results proved in Appendix \ref{appB}. With this relationship between the Lagrangian and Hamiltonian parameters, the formula for the counting of the physical degrees of freedom--known from Dirac's method--can be expressed in terms of the Lagrangian parameters only. With this Lagrangian formula for the counting we can stay in the Lagrangian framework without having to perform the Hamiltonian analysis of the system and to know the physical content of the theory under study from the Lagrangian perspective only. As a bonus, and due to the fact we know the relationship between the Lagrangian and Hamiltonian parameters we can count the number of first- and second-class constraints if we were interested in performing Dirac's analysis.

Section \ref{sec4} contains an illustration of the procedure developed in Sect. \ref{sec3}. The applications include point particle systems and the relativistic free particle. The last example shows how this procedure works in the case where the gauge symmetry is associated with reparametrization invariance, giving the right counting. We also count the physical degrees of freedom of systems that violate the Dirac's conjecture, this is reported in the Appendix \ref{appC}.

Section \ref{sec5} is devoted to uncover the geometrical content of the approach through the covariant canonical formalism. All the examples discussed in the Section \ref{sec4} are also analyzed in this framework  in order to make further clarifications.

Finally, our conclusions are collected in Section \ref{sec6}.

\section{\label{sec2}Algorithm to uncover the Lagrangian gauge symmetries}
In this section we follow the notation and convention of Refs.~\onlinecite{Shirzad,Rothe} and we present a summary of their approach (see also 
Refs. \onlinecite{Teitelboim,Gitman1990} for a generic discussion on the Lagrangian and Hamiltonian formalisms). The relevance--for the present paper--of the material contained in this section is to analyze how the evolution of the variational derivatives is related to the so-called {\it Lagrangian constraints} and the {\it gauge identities}.

So, let us begin with the Lagrangian action principle for a system described with $N$ degrees of freedom $q^i$, $i=1,\ldots, N$, that label the points of the configuration space $\cal{C}$. The motion of the system is described by the curve that makes the action
\begin{equation}
S [q^i]=\int_{t_1}^{t_2}{L(q,\dot{q})dt}, \quad i=1, \ldots, N,\label{1}
\end{equation}
stationary under arbitrary functional variations $\delta q^i(t)$ that vanish at the end points $t_1$ and $t_2$, $\delta q^i (t_1) = 0 = \delta q^i (t_2)$.

The action principle leads to the Euler-Lagrange equations of motion
\begin{equation}\label{falta}
E_{i}^{0}=0, \quad i=1, \ldots, N,
\end{equation}
where
\begin{eqnarray}
E_{i}^{0} &:=& \frac{d}{dt}\left(\frac{\partial L}{\partial \dot{q}^i}\right)-\frac{\partial L}{\partial q^i} \nonumber \\
&\equiv& W_{ij}^{0}(q,\dot{q})\ddot{q}^{j}+K_{i}^{0}(q,\dot{q}),
\label{2}
\end{eqnarray}
is minus the ``variational  derivative''.\cite{Teitelboim} Also
\begin{equation}
W_{ij}^{0}(q,\dot{q}) :=\frac{\partial^{2}L}{\partial\dot{q}^{i}\partial\dot{q}^{j}}, \quad 
K_{i}^{0}(q,\dot{q}) :=\frac{\partial^{2}L}{\partial\dot{q}^{i}\partial{q}^{j}}\dot{q}^{j}-\frac{\partial L}{\partial q^{i}}.
\label{3}
\end{equation}
The superscript ``0'' in $E^0_i$, $W^0_{ij}$, and $K^0_{i}$ is introduced because the form of (\ref{2}) will systematically appear in the approach we are ready to describe and ``0'' means that (\ref{2}) comes just from the definition of the variational derivatives and  there is not time evolution at this stage. The system of equations (\ref{falta}) falls into three categories:
\begin{eqnarray}
(a) && W^0 =0, \quad  K^0 \neq 0, \nonumber\\
(b) && W^0 \neq 0, \quad K^0=0, \nonumber\\
(c) && W^0 \neq 0, \quad K^0 \neq 0,
\end{eqnarray}
where $W^0 =\left ( W^0_{ij}\right )$, $K^0 = \left ( K^0_i \right)$, and $E^0 = \left ( E^0_i \right )$. In the case $(a)$ the system of equations (\ref{falta}) is of first-order in the time derivative.

The singular case is characterized by the vanishing of the determinant of the Hessian matrix $W^0$. Therefore, Rank $W^{0}= N-R'_1$ where $R'_1$ is the number of linearly independent left (right) null  vectors $\lambda_{a_1}(q,\dot{q}) = (\lambda^i_{a_1}) $ of $W^{0}$
\begin{eqnarray}
\lambda_{a_1}^{i}(q,\dot{q})W_{ij}^{0}(q,\dot{q})=0,
\hspace{1cm}
a_1=1,\ldots ,R'_1.
\label{4}
\end{eqnarray}
The left null vectors are linearly independent with the understanding that all $q$'s and $\dot{q}$'s are treated as independent variables.

The procedure we are ready to describe involves a finite number of steps only in the cases $(a)$ and $(c)$  ($K^0 \neq 0$) whereas in the case $(b)$ ($K^0=0$) the procedure begins and ends at step ``0,'' and in this sense is ``trivial.''

Step 0. Contracting (\ref{2}) with each one of the null vectors $\lambda_{a_1}(q,\dot{q})$, we get in the cases $(a)$ and $(c)$
\begin{equation}\label{falta2}
\lambda^i_{a_1} E^0_i =  \lambda_{a_1}^{i}(q,\dot{q})K_{i}^{0}(q,\dot{q}),
\end{equation}
while in the case $(b)$ we have $K^0=0$ from the very beginning and so, instead of (\ref{falta2}), we have
 \begin{eqnarray}\label{trivial}
 \lambda^i_{a_1} E^0_i = 0.
 \end{eqnarray}
Notice that these relations among the $E^0_i$ hold ``off-shell,'' they are named gauge identities. The number of gauge identities is equal to the number of independent null vectors of  $W^0$ (the action for the relativistic free particle analyzed in SubSection \ref{sec4.4} is an example of the case $(b)$ and so of (\ref{trivial})).

More generally, {\it gauge identities} are relations among the $E^0_i$ and their time derivatives only that hold ``off-shell.'' The name comes from the fact that relations of this type are related to the gauge transformations of the $q's$ (this is explained later in this section).

Let us now say some words about relations (\ref{falta2}). Because of the presence of $K_{i}^{0}(q,\dot{q})$, (\ref{falta2}) are not gauge identities generically. Nevertheless, it might happen--depending on the specific form of  $\lambda^i_{a_1}$ and $K_{i}^{0}(q,\dot{q})$--that some combinations of these relations might lead to some gauge identities (an example of this fact is given in SubSection \ref{sec4.2}). Let us denote by $g_0$ the number of gauge identities coming from (\ref{falta2}).

On the other hand,  ``on-shell''  (i.e., assuming that (\ref{falta}) holds) the LHS of (\ref{falta2}) vanishes and this implies
\begin{eqnarray}
\lambda_{a_1}^{i}(q,\dot{q})K_{i}^{0}(q,\dot{q}) =0,
\label{5}
\end{eqnarray}
which are equations among the $q's$ and ${\dot q}'s$ only. They are named {\it Lagrangian constraints}. The number of Lagrangian constraints is equal to the number of independent null vectors of $W^0$, which is $R'_1$. Nevertheless, not all the Lagrangian constraints in (\ref{5}) are functionally independent to each other. Let us denote the independent ones by
\begin{eqnarray}\label{indcons}
\psi_{\bar{a}_1} \left ( q, {\dot q} \right ) =0, \quad \bar{a}_1 =1, \ldots, R'_1 - g_0.
\end{eqnarray}
Therefore, in opposition to gauge identities, Lagrangian constraints hold ``on-shell'' whereas gauge identities hold ``off-shell.'' End of step 0.

Step 1. In the case $(b)$ we do nothing because no Lagrangian constraints arose at step 0. In the cases $(a)$ and $(c)$ we have to handle all the Lagrangian constraints that might have been arisen at step 0. The method demands that all the independent Lagrangian constraints (\ref{indcons}) must be preserved under time evolution, i.e., the time evolution of the independent Lagrangian constraints must vanish too
\begin{eqnarray}
\frac{d}{dt}  \psi_{\bar{a}_1}  =  \frac{\partial \psi_{\bar{a}_1}}{\partial {\dot q}^i} {\ddot q}^i +
\frac{\partial \psi_{\bar{a}_1}}{\partial q^i}{{\dot q}^i} =0.
\end{eqnarray}
These equations must be added to Eqs. (\ref{falta}), and they are arranged as
\begin{eqnarray}
E^1_{i_1} :=
\left (
\begin{array}{c}
  E^0_i   \\
  \frac{d} {dt} \psi_{\bar{a}_1}
\end{array}
\right ) = 0, \quad i_1 =1, \ldots, N+ R'_1 - g_0,
\end{eqnarray}
with
\begin{eqnarray} \label{xxxx}
E^1_{i_1} &=& W^1_{i_1 j} {\ddot q}^j + K^1_{i_1} (q, {\dot q}).
\end{eqnarray}
Obviously, $W^1_{ij} = W^0_{ij}$, $W^1_{N+ {\bar a}_1, j} = \frac{\partial \psi_{\bar{a}_1}}{\partial {\dot q}^j}$, $K^1_{i}=K^0_i$, and $K^1_{N+{\bar a}_1} = \frac{\partial \psi_{\bar{a}_1}}{\partial q^i}{{\dot q}^i}$. Due to the fact $K^1\neq 0$ we are now either in the case $(a)$ or $(c)$ described above. We again apply the same procedure, namely,  we find the left null vectors of $W^1$ and look for the gauge identities and Lagrangian constraints that come from (\ref{xxxx}). In the generic case, at step 1 we get $g_1$ new gauge identities and new Lagrangian constraints, $\psi_{\bar{a}_2} \left ( q, {\dot q} \right )=0$ different from those of the step 0.

Step 2.  We have to add to $E^1_{i_1}=0$ the evolution of the Lagrangian constraints found at step 1, $\frac{d}{dt}  \psi_{\bar{a}_2}=0$,  to build $E^2_{i_2}$ and so on.

The procedure ends at a some finite step either because no new Lagrangian constraints emerge or because only new gauge identities arise. Along the procedure we have to restrict  the analysis to the surface defined by the Lagrangian constraints. This is important because the rank of the matrices $W$'s might change leading to a different set of left null vectors. At the end, we get $g_0 + g_1 + \ldots =g$ gauge identities and $l$ Lagrangian constraints. As we already mentioned, the gauge identities are related to the gauge transformations of the $q$'s. This is explained in what follows.

 The gauge identities, that we already get at the $k$-th step, have the general structure (see Appendix A for its derivation)
\begin{equation}
G_{\mathfrak{g}_k}^{k}:=\sum_{s=0}^{k}\frac{d^s}{dt^s}\left({M^{(k)}}_s^iE_i^{0}\right)=0,
\label{GI}
\end{equation}
where $\mathfrak{g}_k=1,\ldots,g_k$ and ${M^{(k)}}_s^i$ are specific functions of $q$'s and their time derivatives defined by the theory we are dealing with. It is worth noting that time derivatives in (\ref{GI}) are not of arbitrary order and are limited by the step number, $k$, at which the gauge identity belongs. In particular the maximum order that can appear in the whole set of gauge identities is the number of step at which the procedure ends.\\
In this way we exhaust all possible independent relations between variational derivatives and time derivatives of variational derivatives that vanish ``off-shell'' and, because the converse of the Noether's theorem,\cite{Noether} we can find a generating set of gauge transformations in the sense described in Ref.~\onlinecite{Teitelboim}. Let us go into details and multiply (\ref{GI}) by arbitrary functions of time, $\varepsilon^{(k)}$. Using the product rule for derivatives as many times as necessary, we arrive to the expression
\begin{equation}
\sum_{s=0}^{k}\left[(-1)^s\frac{d^s\varepsilon ^{(k)}}{dt^s}{M^{(k)}}^i_s\right] E_i^{0}-\frac{d}{dt}B^{(k)}=0,
\label{20}
\end{equation}
where $B^{(k)}$ is in general a function of $\varepsilon ^{(k)}$, the coordinates and time derivatives thereof. By the converse of Noether construction \cite{Noether} (see also Refs.~\onlinecite{Deriglazov,Arnold}) these are \emph{Noether identities}  and, because of that, (\ref{GI}) sign the gauge invariance of the theory. Therefore the contribution of this step to the gauge transformation at fixed time is
\begin{equation}
\delta_\varepsilon {q^{i(k)}}=\sum_{s=0}^{k}(-1)^s\frac{d^s\varepsilon ^{(k)}}{dt^s}{M^{(k)}}^i_s.
\label{GT}
\end{equation}
Because the iterative procedure described above leads to $g$ gauge identities of the form (\ref{GI}) and for each one we get Noether's identities of the form (\ref{20}), a generating set of the gauge transformation of the Lagrangian action is
\begin{equation}
\delta_\varepsilon q^{i}=\sum_{\alpha =1}^{g}\delta_\varepsilon {q^{i(k_\alpha)}}=\sum_{\alpha =1}^{g}\sum_{s=0}^{k_\alpha}(-1)^s\frac{d^s\varepsilon ^{(k_\alpha)}}{dt^s}{M^{(k_\alpha)}}^i_s\,\,.
\label{17C}
\end{equation}
For each gauge identity appears a relation of the form (\ref{GT}) with an arbitrary function of time $\varepsilon^{(k)}$. Therefore the total number of \emph{arbitrary functions} in the gauge transformation (\ref{17C}) is $g$. This observation will be fundamental in Section \ref{sec3}.

Before ending this section, let us make some remarks:

(1) It was explicitly shown in Ref.~\onlinecite{Shirzad} that the Lagrangian action is invariant under the transformation (\ref{17C}), a fact that is not surprising because of the converse of Noether's theorem.\cite{Noether}

(2) The procedure described above generates gauge transformations at fixed time and from these we can get information for the physical degree of freedom count as we will see later in this paper.

\section{\label{sec3}Physical degree of freedom count}
It was shown in Ref.~\onlinecite{Gracia} that the number of degrees of freedom in the Hamiltonian formalism is that of the Lagrangian formalism.
Therefore we will obtain from the Hamiltonian formalism all the information necessary to make the counting in the Lagrangian formalism and present a closed expression which contains data coming from the Lagrangian procedure only (see also Ref.~\onlinecite{Daniel}).

Let us denote the number of \emph{first-class constraints} by $N_1$, the number of \emph{second-class constraints} by $N_2$ and the number of \emph{primary first-class constraints} by $N_1^{(p)}$.  We will relate the quantities $N_1,N_2,N_1^{(p)}$ with those coming from the current Lagrangian procedure. To get an insight about it, let us recall that in the Hamiltonian and the Lagrangian approaches the number of primary Hamiltonian constraints and the number of primary Lagrangian constraints plus gauge identities at level zero, respectively, are equal to Rank $W^0$. As a next step in both procedures, we evolve the constraints until no new information emerges. Then, it is natural to think that the number of relations we get in both processes are equal. Thus, the total number of Hamiltonian constraints must be equal to the total number of Lagrangian constraints $l$ plus the number of gauge identities $g$, i.e.,
\begin{eqnarray}\label{key1}
N_1+N_2=l+g.
\end{eqnarray}
As another fact, the gauge invariance of the Lagrangian theory is represented by the gauge parameters $\varepsilon$'s. Since at any given time we can choose arbitrarily any value for the gauge parameters and its time derivatives, it is pointed out that the gauge generators of the Hamiltonian theory (\emph{the first-class constraints}) must be related to the gauge parameters and its time derivatives. In fact, if the total number of gauge parameters plus its successive derivatives is denoted by $e$, we have that
\begin{eqnarray}\label{key2}
N_1 = e.
\end{eqnarray}
Combining (\ref{key1})  and (\ref{key2}), we get
\begin{eqnarray}\label{key3}
N_2=l+g-e,
\end{eqnarray}
and the classical Hamiltonian formula for the physical degree of freedom count becomes
\begin{eqnarray}\label{I1}
&=& N - N_1 - \frac12 N_2 \nonumber\\
&=& N-e-\frac{1}{2}\left(l+g-e\right) \nonumber \\
&\equiv& N-\frac{1}{2}\left(l+g+e\right).
\end{eqnarray}

In order to proof the previous insight, we need the following two results:

{\bf Result 1}.  {The total number of arbitrary independent functions of time $(g)$ appearing in the Lagrangian form of the gauge transformation law is equal to the number of first-class primary constraints, i.e, 
\begin{eqnarray}\label{result1}
N_1^{(p)}=g.
\end{eqnarray}

{\bf Result 2}. The total number of ``effective'' gauge parameters $(e)$ is equal to the total number of first-class constraints, where by ``effective'' it is understood that we count independently the gauge parameters and its successive time derivatives, i.e, 
\begin{eqnarray}\label{result2}
N_1=e.
\end{eqnarray}

The proof of the results 1 and 2 can be found in the Appendix \ref{appB}. The proof is close in form to the one reported in Refs.~\onlinecite{Zanelli, Teitelboim} for first-class systems but now it is generalized to include all kind of systems.

To continue, we recall a result contained in the theorem 2 of Ref.~\onlinecite{Pons}, which states that if $l$ is the total number of independent Lagrangian constraints and the total number of independent Hamiltonian constraints is $N_1+N_2$, then
\begin{equation}
l=N_1+N_2-N_1^{(p)}.
\end{equation}
Combining this fact with (\ref{result1}) and (\ref{result2}),  we obtain
\begin{eqnarray}
N_2 =l+N_1^{(p)}-N_1\equiv l+g-e.
\label{28}
\end{eqnarray}
Finally, the number of physical degrees of freedom is
\begin{eqnarray}
N-N_1-\frac{1}{2}N_2=N-e-\frac{1}{2}\left(l+g-e\right) = N-\frac{1}{2}\left(l+g+e\right),
\label{29}
\end{eqnarray}
in agreement with the insight (\ref{I1}).

\section{\label{sec4}Examples}
We now illustrate the theoretical framework developed in Section \ref{sec3}.
\subsection{An example of the case (a)}\label{sec4.1}
Let us consider the system reported in Ref.~\onlinecite{Chitaia} and defined by the Lagrangian
\begin{equation}
L(q^i,\dot{q}^i)=\dot{q}^1 q^2 - \dot{q}^2 q^1- (q^1 - q^ 2) q^3.
\label{30}
\end{equation}
\subsubsection{Getting the gauge identities and the Lagrangian constraints:}
We have
\begin{eqnarray}\label{ejemploa}
E^0 = \left (
\begin{array}{r}
2\dot{q}^2 + q^3\\
-2\dot{q}^1- q^3\\
q^1 - q^2
\end{array}
\right ) = W^0 \left (
\begin{array}{c}
{\ddot q}^1 \\ {\ddot q}^2 \\ {\ddot q}^3
\end{array}
\right )+ K^0 = K^0.
\end{eqnarray}
Because of $W^0=0$ we are in the case $(a)$ and the left null vectors of $W^0$ are trivial, i.e.,
\begin{eqnarray}
\lambda_1 = \left ( 1,0,0 \right ),\quad
\lambda_2 = \left ( 0,1,0 \right ), \quad
\lambda_3 = \left ( 0,0,1 \right ).\label{31b}
\end{eqnarray}
Therefore, the contraction of (\ref{ejemploa}) with the null vectors (\ref{31b}) gives
\begin{eqnarray}
E^0 = K^0,
\label{32}
\end{eqnarray}
which is the original relation (\ref{ejemploa}). These relations are independent among themselves and we do not have gauge identities step 0.

Now, from (\ref{32}) and $E^0 =0$ we get
\begin{eqnarray}
K^0=0.
\end{eqnarray}
Because the components of $K^0$ are independent among themselves, we get the constraints
\begin{eqnarray} \label{primerascons}
\psi := K^0 =0.
\end{eqnarray}
Therefore, at step 0 we get no gauge identities, $g_0=0$, and three constraints (\ref{primerascons}), $l_0=3$.

Step 1. The constraints (\ref{primerascons}) must be preserved under time evolution, so we have to add ${\dot \psi}=0$ to the equations of motion $E^0=0$, and again look for the left null vectors
\begin{equation}
E^{1}=\left(
\begin{array}{c}
E^{0} \\
\frac{d}{dt} \psi
\end{array}
\right)=\left(
\begin{array}{r}
2\dot{q}^2 + q^3\\
-2\dot{q}^1-q^3\\
q^1- q^2\\
2\ddot{q}^2+\dot{q}^3\\
-2\ddot{q}^1-\dot{q}^3\\
\dot{q}^1 - \dot{q}^2
\end{array}\right),
\label{33}
\end{equation}
and rewrite it as
\begin{eqnarray}
E^1 &=& W^1 \left (
\begin{array}{c}
{\ddot q}^1 \\ {\ddot q}^2 \\ {\ddot q}^3
\end{array}
\right ) + K^1 \nonumber\\
&=& \left(
\begin{array}{l c r}
0 & 0 & 0\\
0 & 0 & 0\\
0 & 0 & 0\\
0 & 2 & 0\\
-2 & 0 & 0\\
0 & 0 & 0
\end{array}
\right)   \left (
\begin{array}{c}
{\ddot q}^1 \\ {\ddot q}^2 \\ {\ddot q}^3
\end{array}
\right ) + \left(
\begin{array}{c}
2\dot{q}^2+q^3\\
-2\dot{q}^1- q^3\\
q^1-q^2\\
\dot{q}^3\\
-\dot{q}^3\\
\dot{q}^1-\dot{q}^2
\end{array}
\right).
\end{eqnarray}
The left null vectors of $W^1$ are
\begin{eqnarray}
\lambda_1 &=& \left (1,0,0,0,0,0 \right ), \quad \lambda_2 = \left ( 0,1,0,0,0,0 \right ), \nonumber\\
\lambda_3 &=& \left (0,0,1,0,0,0 \right ), \quad \lambda_4 = \left ( 0,0,0,0,0,1 \right ).
\end{eqnarray}
The first three vectors are those of (\ref{31b}) but augmented by 3 zeros and the contraction of $E^1$ with them gives again (\ref{32}). $\lambda_4$ is a new left null vector and its contraction with $E^{1}$ gives
\begin{equation}
\frac{d}{dt} \psi_3 =\dot{q}^1- \dot{q}^2,
\label{34}
\end{equation}
where $\psi_3$ is the third component of $\psi$. Now, using (\ref{32}) and (\ref{primerascons}), we rewrite $\psi_3$ as $\psi_3 =E^0_3$ and the RHS of (\ref{34}) as ${\dot q}^1 - {\dot q}^2 = - \frac12 E^0_2 - \frac12 E^0_1$ and therefore (\ref{34}) acquires the form
\begin{equation}
G_1^{1} :=\frac{d}{dt}E_3^{0}+\frac{1}{2}E_1^{0}+\frac{1}{2}E_2^{0}=0.
\label{35}
\end{equation}
Because (\ref{35}) holds \emph{off-shell}, it corresponds to a genuine \emph{gauge identity}. Therefore, the method stops here, at step 1, which involves one gauge identity (\ref{35}), $g_1=1$, and no Lagrangian constraints, $l_1=0$.

\subsubsection{Local gauge symmetries:}
The gauge symmetries come from the gauge identities. Therefore, multiplying (\ref{35}) by an arbitrary function of
time $\varepsilon$ and rewriting it, we get the Noether's identity
\begin{equation}
\frac{\varepsilon}{2}E_1^{0}+\frac{\varepsilon}{2}E_2^{0}-\dot{\varepsilon}E_3^{0}+\frac{d}{dt}\left(\varepsilon E_3^{0}\right)=0,
\label{36}
\end{equation}
and, from inspection, we can directly read the gauge transformation
\begin{equation}
\delta_\varepsilon q^1=\frac{\varepsilon}{2},
\hspace{1cm}
\delta_\varepsilon q^2=\frac{\varepsilon}{2},
\hspace{1cm}
\delta_\varepsilon q^3=-\dot{\varepsilon},
\label{37}
\end{equation}
which involves the parameter $\varepsilon$ as well as its time derivative ${\dot\varepsilon}$, i.e. two effective gauge parameters.

\subsubsection{Degree of freedom count:}
In summary, we have $l= l_0 + l_1 = 3 + 0 =3$ independent Lagrangian constraints, $g = g_0 + g_1 = 0 + 1 =1$ gauge identities, and $e=2$ effective gauge parameters (${\varepsilon}$ and ${\dot\varepsilon}$). Using the expression (\ref{29}), the number of physical degrees of freedom is
\begin{equation}
3-\frac{1}{2}(3+1+2)=0.
\label{38}
\end{equation}
Furthermore, we can get information about the Hamiltonian analysis {\it without} having to perform it. For instance, from (\ref{key2}) and (\ref{key3}) the number of first-class constraints is $N_1=e \equiv 2$ and the number of second-class constraints is $N_2=l+g-e\equiv 2$, which is in agreement with the Hamiltonian analysis reported in Ref.~\onlinecite{Chitaia}.

\subsection{An example where gauge identities emerge from the very beginning}\label{sec4.2}
Let us start with the Lagrangian
\begin{equation}
L(q^i,\dot{q}^i)=(\dot{q}^1+\dot{q}^2) q^3+\frac{1}{2} \left ( \dot{q}^3 \right)^2.
\label{LH1}
\end{equation}
\subsubsection{Lagrangian analysis:}
We have $(q^i)=(q^1,q^2,q^3)$ as coordinates for the configuration space. The variational derivatives associated with (\ref{LH1}) are
\begin{equation}
E^0=\left(
\begin{array}{c}
\dot{q}^3\\
\dot{q}^3\\
\ddot{q}^3- \dot{q}^1- \dot{q}^2
\end{array}
\right)
\equiv W^0\left(
\begin{array}{c}
\ddot{q}^1\\
\ddot{q}^2\\
\ddot{q}^3
\end{array}
\right)+K^0,
\label{LH2}
\end{equation}
with
\begin{equation}
W^0=\left(
\begin{array}{c c c}
0 & 0 & 0\\
0 & 0 & 0\\
0 & 0 & 1
\end{array}
\right),\hspace{5mm}
K^0=\left(
\begin{array}{c}
\dot{q}^3\\
\dot{q}^3\\
-\dot{q}^1-\dot{q}^2
\end{array}
\right).
\label{LH3}
\end{equation}
A basis for the left null vectors of $W^0$ is given by
\begin{eqnarray}
\lambda_1 = \left (1,0,0 \right ), \quad \lambda_2 = \left (0,1,0 \right ). \label{LH4}
\end{eqnarray}
The contraction of (\ref{LH2}) with the null vectors leads to
\begin{equation}
E_1^0=\dot{q}^3, \hspace{6mm}
E_2^0=\dot{q}^3.
\label{LH5}
\end{equation}
Even though $\lambda_1$ and $\lambda_2$ are linearly independent,  the contractions in (\ref{LH5}) are not functionally independent. From (\ref{LH5}) we get
\begin{equation}
G^0:=E_1^0-E_2^0=0,
\label{LH7}
\end{equation}
which holds off-shell and thus it corresponds to a genuine gauge identity.

Now, from (\ref{LH5}) and $E^0=0$ (on-shell) we get
\begin{eqnarray}
{\dot q}^3 =0, \quad {\dot q}^3=0,
\end{eqnarray}
implying just one independent constraint
\begin{eqnarray}\label{lc777}
\psi_1 := {\dot q}^3 =0.
\end{eqnarray}
End of step 0, which involves $g_0=1$ gauge identity (\ref{LH7}) and $l_0 =1$ Lagrangian constraint (\ref{lc777}).

Step 1. The constraint (\ref{lc777}) must be preserved under time evolution, ${\dot \psi}_1 =0$, and we have to add this equation to the equations of motion $E=0$. By doing this, we get
\begin{eqnarray}
E^{1}=\left(
\begin{array}{c}
E^{0}\\
\frac{d}{dt} \psi_1
\end{array}
\right) & =\left(
\begin{array}{c}
\dot{q}^3\\
\dot{q}^3\\
\ddot{q}^3 - \dot{q}^1-\dot{q}^2\\
\ddot{q}^3
\end{array}\right)\\
        & = W^1\left(\begin{array}{c}
        \ddot{q}^1\\
        \ddot{q}^2\\
        \ddot{q}^3
        \end{array}
        \right)+K^1,
\label{LH10}
\end{eqnarray}
with
\begin{equation}
W^1=\left(
\begin{array}{c c c}
0 & 0 & 0\\
0 & 0 & 0\\
0 & 0 & 1\\
0 & 0 & 1
\end{array}
\right),\hspace{5mm}
K^1=\left(
\begin{array}{c}
\dot{q}^3\\
\dot{q}^3\\
-\dot{q}^1-\dot{q}^2\\
0
\end{array}
\right).
\label{LH11}
\end{equation}
A basis for the left null vectors of $W^1$ is
\begin{eqnarray}
\lambda_1 = (1,0,0,0), \quad \lambda_2 = (0,1,0,0), \quad \lambda_3 = (0,0,-1,1). \label{LH12}
\end{eqnarray}
The first two vectors are those of (\ref{LH4}) augmented by one zero and their contraction with $E^1$ gives again (\ref{LH5}). On the
other hand, $\lambda_3$ is a new left null vector and its contraction
with $E^1$ gives
\begin{equation}
- E^0_3 + \frac{d}{dt} \psi_1 =\dot{q}^1+\dot{q}^2.
\label{LH13}
\end{equation}
This relation is not a gauge identity. Therefore, imposing $E^1=0$ (not just $E^0$), we get the Lagrangian constraint
\begin{eqnarray}\label{otra+}
\psi_2 := {\dot q}^1 + {\dot q}^2 =0.
\end{eqnarray}
End of step 1, which includes $g_1=0$ gauge identities and $l_1=1$ Lagrangian constraint (\ref{otra+}).

Step 2. We have to add ${\dot \psi}_2=0$ to $E^1=0$
\begin{eqnarray}
E^{2}=\left(
\begin{array}{c}
E^0 \\
\frac{d}{dt}\psi_1\\
\frac{d}{dt}\psi_2
\end{array}
\right) & =\left(
\begin{array}{c}
\dot{q}^3\\
\dot{q}^3\\
\ddot{q}^3 - \dot{q}^1 - \dot{q}^2\\
\ddot{q}^3\\
\ddot{q}^1+\ddot{q}^2
\end{array}\right)\\
        & = W^2\left(\begin{array}{c}
        \ddot{q}^1\\
        \ddot{q}^2\\
        \ddot{q}^3
        \end{array}
        \right)+K^2,
\label{LH14}
\end{eqnarray}
where
\begin{equation}
W^2=\left(
\begin{array}{c c c}
0 & 0 & 0\\
0 & 0 & 0\\
0 & 0 & 1\\
0 & 0 & 1\\
1 & 1 & 0
\end{array}
\right),\hspace{5mm}
K^2=\left(
\begin{array}{c}
\dot{q}^3\\
\dot{q}^3\\
-\dot{q}^1-\dot{q}^2\\
0\\
0
\end{array}
\right).
\label{LH15}
\end{equation}
A basis for the left null vectors of $W^2$
\begin{eqnarray}
\lambda_1 = \left (1,0,0,0,0 \right ), \quad \lambda_2 = \left ( 0,1,0,0,0 \right ), \quad 
\lambda_3 = \left ( 0,0,-1,1,0 \right ). \label{LH16}
\end{eqnarray}
All vectors are those of (\ref{LH12})  augmented by zeros and their contraction with $E^2$ gives again (\ref{LH5}) and (\ref{LH13}). Therefore, the procedure ends at step 2, which includes $g_2=0$ gauge identities and $l_2=0$ Lagrangian constraints.

Now we use the gauge identities to get the gauge transformations. The product of (\ref{LH7}) with an arbitrary function of time $\varepsilon(t)$ gives the Noether's identity
\begin{equation}
\varepsilon E_1^0-\varepsilon E_2^0=0.
\label{LH8}
\end{equation}
By inspection, we can read the gauge transformation
\begin{equation}
\delta_\varepsilon q^1 =\varepsilon,
\hspace{1cm}
\delta_\varepsilon q^2 =-\varepsilon,
\hspace{1cm}
\delta_\varepsilon q^3 =0.
\label{LH9}
\end{equation}
Notice that just $\varepsilon$ (and not $\dot{\varepsilon}$) is involved in the gauge transformation, i.e., there is one effective gauge parameter $\varepsilon$ .

In summary, we have 2 independent Lagrangian constraints ($l=l_0 + l_1 + l_2= 1+ 1+0 = 2$), 1 gauge identity ($g=g_0 +g_1+g_2=1+0+0=1$) and $e=1$ effective gauge parameter. Using the expression (\ref{29}), the number of physical degrees of freedom is
\begin{equation}
3-\frac{1}{2}(2+1+1)=1.
\label{LH18}
\end{equation}
Moreover ({\ref{key2}) and (\ref{key3}) indicate that in the Hamiltonian analysis we must get $N_1=e \equiv 1$ first-class constraints and $N_2=l+g-e\equiv 2$ second-class constraints. This is indeed so and it is explained in what follows.
\subsubsection{Hamiltonian analysis:}
Dirac's method calls for the definition of the momenta $(p_i)$ canonically conjugate to the configuration variables ($q^i$). Therefore, from (\ref{LH1})
\begin{equation}
p_1:=\frac{\partial L}{\partial\dot{q}^1}=q^3,\hspace{5mm}
p_2:=\frac{\partial L}{\partial\dot{q}^2}=q^3,\hspace{5mm}
p_3:=\frac{\partial L}{\partial\dot{q}^3}=\dot{q}^3,
\label{LH19}
\end{equation}
which  imply the primary constraints
\begin{equation}
\phi_1=p_1- q^3\approx 0,\hspace{5mm}
\phi_2=p_2-q^3\approx 0.
\label{LH20}
\end{equation}
The primary Hamiltonian is $H_0=\frac{1}{2}p_3^2$ and then the total Hamiltonian is
\begin{equation}
H_T=H_0+\mu_1\phi_1+\mu_2\phi_2,
\label{LH21}
\end{equation}
where $\mu_1$ and $\mu_2$ are Lagrange multipliers. The evolution of the primary constraints, $\dot{\phi}_1=\left\{\phi_1,H_T\right\}\approx 0$ and $\dot{\phi}_2=\left\{\phi_2,H_T\right\}\approx 0$, implies the secondary constraint
\begin{equation}
\chi:=-p_3\approx 0.
\label{LH22}
\end{equation}
The evolution of the secondary constraints, $\dot{\chi}=\left\{\chi,H_T\right\}\approx 0$, implies the relation between the Lagrange multipliers $-\mu_1-\mu_2\approx 0$. Therefore, no more constraints arise and Dirac's procedure calls for the classification of the constraints. It turns out that
$\Omega^{(1)}=-\phi_1+\phi_2$ is first-class and that there are two second-class constraints, for instance $\phi_1$ and $\chi$, in fully agreement with the prediction of the previous Lagrangian analysis. Furthermore, $H_0$ is first-class because $\chi$ is second-class.
\subsection{An example involving a high order gauge transformation}\label{sec4.3}
We would like to illustrate the procedure in a theory of first order with gauge transformations of second order. Let us consider the system described by the first order Lagrangian \cite{Zanelli}
\begin{equation}
L=\frac{1}{2}\left[\left(\dot{q}^2-e^{q^1}\right)^2+\left(\dot{q}^3-q^2\right)^2\right].
\label{H4}
\end{equation}
The variational derivatives associated with (\ref{H4}) are
\begin{subequations}
\begin{eqnarray}
E_1^0 & := & \frac{d}{dt}\left(\frac{\partial L}{\partial\dot{q}^1}\right)-\frac{\partial L}{\partial q^1}=e^{q^1}\left(\dot{q}^2-e^{q^1}\right),\\
E_2^0 & := & \frac{d}{dt}\left(\frac{\partial L}{\partial\dot{q}^2}\right)-\frac{\partial L}{\partial q^2}=\ddot{q}^2-\dot{q}^1e^{q^1}+\dot{q}^3-q^2,\\
E_3^0 & := & \frac{d}{dt}\left(\frac{\partial L}{\partial\dot{q}^3}\right)-\frac{\partial L}{\partial q^3}=\ddot{q}^3-\dot{q}^2.
\end{eqnarray}
\label{H5}
\end{subequations}
Step 0: We can rewrite (\ref{H5}) as
\begin{equation}
E^0=W^0\left(
\begin{array}{c}
\ddot{q}^1\\
\ddot{q}^2\\
\ddot{q}^3
\end{array}
\right)+K^0,
\label{H6}
\end{equation}
with
\begin{equation}
W^0=\left(
\begin{array}{c c c}
0 & 0 & 0\\
0 & 1 & 0\\
0 & 0 & 1
\end{array}
\right),\hspace{1mm}
K^0=\left(
\begin{array}{c}
e^{q^1}\left(\dot{q}^2-e^{q^1}\right)\\
-\dot{q}^1e^{q^1}+\dot{q}^3-q^2\\
-\dot{q}^2
\end{array}
\right).
\label{H7}
\end{equation}
A basis for the left null vectors of $W^0$ is given by
\begin{equation}
\lambda_1=\left(1,0,0\right).
\label{H8}
\end{equation}
The contraction of (\ref{H6}) with the null vector leads to
\begin{equation}
E^0_1=e^{q^1}\left(\dot{q}^2-e^{q^1}\right),
\label{H9}
\end{equation}
which ``on-shell"($E^0=0$) turns out to be the Lagrangian constraint
\begin{equation}
\psi_1:=e^{q^1}\left(\dot{q}^2-e^{q^1}\right)=0.
\label{H10}
\end{equation}
End of step 0, which involves $g_0=0$ gauge identities and $l_0=1$ Lagrangian constraint (\ref{H10}).

Step 1: The constraint (\ref{H10}) must be preserved under time evolution, $\dot{\psi}_1=0$, and we have to add this equation to the equation of motion $E^0=0$. By doing this we get
\begin{equation}
E^1=\left(
\begin{array}{c}
E^0\\
\frac{d}{dt}\psi_1
\end{array}
\right)\equiv W^1\left(
\begin{array}{c}
\ddot{q}^1\\
\ddot{q}^2\\
\ddot{q}^3
\end{array}
\right)+K^1,
\label{H11}
\end{equation}
where
\begin{equation}
W^1=\left(
\begin{array}{c c c}
0 & 0 & 0\\
0 & 1 & 0\\
0 & 0 & 1\\
0 & e^{q^1} & 0
\end{array}
\right),\quad
K^1=\left(
\begin{array}{c}
e^{q^1}\left(\dot{q}^2-e^{q^1}\right)\\
-\dot{q}^1e^{q^1}+\dot{q}^3-q^2\\
-\dot{q}^2\\
e^{q^1}\left(\dot{q}^1\dot{q}^2-2\dot{q}^1e^{q^1}\right)
\end{array}
\right).
\label{H12}
\end{equation}
A basis for the left null vectors of $W^1$ is
\begin{equation}
\lambda_1=\left(1,0,0,0\right),\hspace{5mm}
\lambda_2=\left(0,-1,0,e^{-q^1}\right).
\label{H13}
\end{equation}
The first vector is that of (\ref{H8}) augmented by one zero and its contraction with $E^1$ gives again (\ref{H9}). On the other hand, $\lambda_2$ is a new left null vector and its contraction with $E^1$ gives
\begin{equation}
-E^0_2+e^{-q^1}\frac{d}{dt}\psi_1=\dot{q}^1\left(\dot{q}^2-e^{q^1}\right)-\dot{q}^3+q^2.
\label{H14}
\end{equation}
This relation is not a gauge identity. Therefore, imposing $E^1=0$ (not just $E^0=0$) and $\psi_1=0$ from (\ref{H10}), we get the Lagrangian constraint
\begin{equation}
\psi_2:=-\dot{q}^3+q^2=0.
\label{H15}
\end{equation}
End of step 1, which includes $g_1=0$ gauge identities and $l_1=1$ Lagrangian constraint (\ref{H15}).

Step 2: We have to add $\dot{\psi}_2=0$ to $E^1=0$
\begin{equation}
E^2=\left(
\begin{array}{c}
E^0\\
\frac{d}{dt}\psi_1\\
\frac{d}{dt}\psi_2
\end{array}
\right)\equiv W^2\left(
\begin{array}{c}
\ddot{q}^1\\
\ddot{q}^2\\
\ddot{q}^3
\end{array}
\right)+K^2,
\label{H16}
\end{equation}
where
\begin{equation}
W^2=\left(
\begin{array}{c c c}
0 & 0 & 0\\
0 & 1 & 0\\
0 & 0 & 1\\
0 & e^{q^1} & 0\\
0 & 0 & -1
\end{array}
\right),\quad
K^2=\left(
\begin{array}{c}
e^{q^1}\left(\dot{q}^2-e^{q^1}\right)\\
-\dot{q}^1e^{q^1}+\dot{q}^3-q^2\\
-\dot{q}^2\\
e^{q^1}\left(\dot{q}^1\dot{q}^2-2\dot{q}^1e^{q^1}\right)\\
\dot{q}^2
\end{array}
\right).
\label{H17}
\end{equation}
A basis for the left null vectors of $W^2$
\begin{eqnarray}\label{H18}
\lambda_1  = \left(1,0,0,0,0\right),\quad  \lambda_2=\left(0,-1,0,e^{-q^1},0\right), \quad
\lambda_3  = \left(0,0,-1,0,-1\right).
\end{eqnarray}
The first two vectors are those of (\ref{H13}) augmented by one zero and their contraction with $E^2$ gives again (\ref{H9}) and (\ref{H14}). On the other hand, $\lambda_3$ is a new left null vector and its contraction with $E^2$ gives
\begin{equation}
-E_3^0-\frac{d}{dt}\psi_2=0.
\label{H21}
\end{equation}
Using (\ref{H9}) and (\ref{H10}), we rewrite $\psi_1$ as $\psi_1=E^0_1$. Additionally from (\ref{H14}), (\ref{H15}) and (\ref{H9}) we can rewrite $\psi_2$ as $\psi_2=-E^0_2+e^{-q^1}\frac{d}{dt}E^0_1-\dot{q}^1e^{-q^1}E^0_1$. Therefore (\ref{H21}) acquires the form
\begin{equation}
G^2:=E_3^0+\frac{d}{dt}\left(-E_2^0+e^{-q^1}\frac{d}{dt}E_1^0-\dot{q}^1e^{-q^1}E^0_1\right)=0.
\label{H23}
\end{equation}
Rewriting it leads to
\begin{eqnarray}
G^2:= & e^{-q^1}\frac{d^2}{dt^2}E_1^0-2\dot{q}^1e^{-q^1}\frac{d}{dt}E_1^0-\frac{d}{dt}E_2^0 \nonumber \\
& -\ddot{q}^1e^{-q^1}E_1^0+(\dot{q}^1)^2e^{-q^1}E_1^0+E_3^0=0. \label{H24}
\end{eqnarray}

Because (\ref{H24}) holds \emph{off-shell}, it corresponds to a genuine \emph{gauge identity}. Notice (\ref{H24}) has the general structure (\ref{GI}) depicted in Section \ref{sec2}. Therefore the method ends here, at step 2, which involves one gauge identity (\ref{H24}), $g_2=1$, and no Lagrangian constraints, $l_2=0$.\\
Now we use the gauge identity to get the gauge transformation. Multiplying (\ref{H24}) with an arbitrary function of time $\varepsilon(t)$ and rewriting the resulting expression gives the Noether's identity
\begin{eqnarray}
\left(e^{-q^1}\ddot{\varepsilon} \right)E^0_1 & +\left(\dot{\varepsilon}\right)E^0_2+\left(\varepsilon\right)E^0_3-\frac{d}{dt}\left[\varepsilon\dot{q}^1 e^{-q^1}E^0_1+\varepsilon E^0_2\right. \nonumber \\
& \left. +\dot{\varepsilon} e^{-q^1}E^0_1-\varepsilon e^{-q^1}\frac{d}{dt}E^0_1\right]=0. \label{H25}
\end{eqnarray}
Notice also that (\ref{H25}) has the structure (\ref{20}) depicted in Section \ref{sec2}. By inspection, we can read the gauge transformation
\begin{equation}
\delta_\varepsilon q^1=e^{-q^1}\ddot{\varepsilon},\hspace{5mm}
\delta_\varepsilon q^2=\dot{\varepsilon},\hspace{5mm}
\delta_\varepsilon q^3=\varepsilon.
\label{H26}
\end{equation}
Notice that $\varepsilon$, $\dot{\varepsilon}$ and $\ddot{\varepsilon}$ are involved in the gauge transformation. Therefore the number of effective parameters is $e=3$.\\
Finally, we have 2 independent Lagrangian constraints ($l=l_0+l_1+l_2=1+1+0=2$), 1 gauge identity ($g=g_0+g_1+g_2=0+0+1=1$) and $e=3$ effective parameters. Using expression (\ref{29}), the number of physical degrees of freedom is
\begin{equation}
3-\frac{1}{2}\left(2+1+3\right)=0.
\label{H27}
\end{equation}
Moreover, (\ref{key2}) and (\ref{key3}) indicate that in the Hamiltonian analysis we must get $N_1=e=3$ first-class constraints and $N_2=l+g-e=0$ second-class constraints, a fact that will be verified below.\\
It is worth noting that the order of the gauge identity always is less or equal to the numbers of steps in the Lagrangian analysis and so it is not arbitrary.

The variation of the Lagrangian (\ref{H4}) under transformations (\ref{H26}) is
\begin{equation}
\delta_\varepsilon L=0.
\label{H28}
\end{equation}

Finally, it is worth mentioning that the gauge identity (\ref{H24}) agrees with the one found in Ref. \onlinecite{Kiriushcheva2}. Nevertheless, in  Ref. \onlinecite{Kiriushcheva2} it was obtained from the knowledge of the gauge transformation of the Hamiltonian analysis. Here, in opposition, we generate this gauge identity from the Lagrangian formalism only and then we find the gauge transformation (\ref{H26}), avoiding the Hamiltonian analysis.

\emph{Hamiltonian analysis}\\
For the sake of completeness, and in order to compare the previous Lagrangian approach with the Dirac's Hamiltonian formalism, we review here the Hamiltonian analysis of the Lagrangian (\ref{H4}), which was originally reported in Ref. \onlinecite{Zanelli}.

Dirac's method calls for the definition of the momenta $(p_i)$ canonically conjugate to the configuration variables ($q^i$). Therefore, from (\ref{H4})
\begin{equation}
p_1:=\frac{\partial L}{\partial\dot{q}^1}=0,\quad
p_2:=\frac{\partial L}{\partial\dot{q}^2}=\dot{q}^2-e^{q^1},\quad
p_3:=\frac{\partial L}{\partial\dot{q}^3}=\dot{q}^3-q^2,
\label{H29}
\end{equation}
which  imply the primary constraint
\begin{equation}
\phi_1=p_1\approx 0.
\label{H30}
\end{equation}
The primary Hamiltonian is
\begin{equation}
H_0=\frac{1}{2}\left(p_2^2+p_3^2\right)+e^{q^1}p_2+q^2p_3,
\label{H31}
\end{equation}
and then the total Hamiltonian is
\begin{equation}
H_T=H_0+\mu\phi_1,
\label{H32}
\end{equation}
where $\mu$ is a Lagrange multiplier. The evolution of the primary constraint, $\dot{\phi}_1=\left\{\phi_1,H_T\right\}\approx 0$, implies the secondary constraint
\begin{equation}
\phi_2:=e^{q^1}p_2\approx 0,
\label{H33}
\end{equation}
and the evolution of the secondary constraint, $\dot{\phi}_2=\left\{\phi_2,H_T\right\}\approx 0$, implies a new secondary constraint
\begin{equation}
\phi_3:=e^{q^1}p_3\approx 0.
\label{H34}
\end{equation}
Finally, the evolution of $\phi_3$ does not generate a new constraint and the method ends here.\\
Now, Dirac's procedure calls for the classification of the constraints. It turns out that the nontrivial components of the constraint algebra are
\begin{equation}
\{\phi_1,\phi_2\}=-\phi_2,\quad
\{\phi_1,\phi_3\}=-\phi_3, \quad
\{\phi_2,\phi_3\}=0.
\label{H35}
\end{equation}
Therefore, all constraints are first-class ($N_1=3$, $N_2=0$), in fully agreement with the prediction of the previous Lagrangian analysis. Furthermore,
\begin{equation}
H_0=\frac{1}{2}\left(e^{-2q^1}\phi_2^2+e^{-2q^1}\phi_3^2\right)+\phi_2+e^{-q^1}q^2\phi_3
\label{H36}
\end{equation}
is first-class because it is a combination of first-class quantities.

\subsection{An example of the case (b): the relativistic free particle}\label{sec4.4}
A Lagrangian action that describes the motion of a relativistic free particle in Minkowski space-time $(\mathbb{R}^4, \eta)$ is
\begin{equation}
S[x^\mu ]=-mc\int_{r_1}^{r_2}\sqrt{-\eta_{\mu\nu}\frac{dx^\mu(r)}{dr}\frac{dx^\nu(r)}{dr}}dr, \label{rfpa}
\end{equation}
where $r$ is the evolution parameter, $\eta= \eta_{\mu\nu} d x^{\mu} d x^{\nu}$ is the Minkowski metric with $(\eta_{\mu\nu})=\mbox{diag} (-1,1,1,1)$, $x^{\mu}$ are Minkowski coordinates. This action is generally covariant under reparametrizations of the world line that do not change its orientation, namely
\begin{eqnarray}\label{rfpgauges}
r' &=& f(r), \quad {x'}^{\mu}(r') = x^{\mu} (r),
\end{eqnarray}
with $\frac{d f(r)}{dr} >0$.

In this case, we have
\begin{eqnarray}\label{otra}
E^0_\alpha  &=& mc \left\{\frac{- \left ( {\dot x}^{\beta} {\dot x}_{\beta} \right ) {\ddot x}_{\alpha} + {\dot x}_{\alpha} \left ( {\dot x}_{\beta} {\ddot x}^{\beta}\right )}{(-\eta_{\mu\nu}\dot{x}^\mu\dot{x}^\nu)^{3/2}}\right\}\nonumber\\
		 &=& W^0_{\alpha\beta} {\ddot x}^{\beta} + K^0_{\alpha},
\end{eqnarray}
with
\begin{eqnarray}
W^0_{\alpha\beta} &=& \frac{mc}{\left ( - \eta_{\mu\nu} {\dot x}^{\mu} {\dot x}^{\nu} \right )^{3/2}} \left ( - \eta_{\mu\nu} {\dot x}^{\mu} {\dot x}^{\nu} \eta_{\alpha\beta} + {\dot x}_{\alpha} {\dot x}_{\beta} \right ), \nonumber\\
K^0_{\alpha} &=& 0.
\end{eqnarray}
Therefore, this means that we are in the case $(b)$. The Hessian matrix $W^0$ has just one left null vector given by
\begin{eqnarray}
\lambda = ({\dot x}^{\beta})
\end{eqnarray}
The contraction of $\lambda$ with (\ref{otra}) leads to the gauge identity
\begin{equation}\label{rfpgauge}
G^0 := \dot{x}^\alpha E^0_\alpha=0.
\end{equation}
As we already explained, there is no way to get Lagrangian constraints in the case $(b)$. The more we can get is gauge identities and we already have all of them. Therefore, the procedure stops here, at step ``0''. It is expected that the gauge identity (\ref{rfpgauge}) is related to the gauge symmetry of the action (\ref{rfpgauges}) and, in fact, this is so by multiplying this equation with $- \delta r$ to get
\begin{equation}
 \left ( -\dot{x}^\alpha \delta r \right ) E_\alpha=0,
\end{equation}
from which,
\begin{equation}
\delta_{\mbox{gauge}} x^{\alpha} = - {\dot x}^{\alpha} \delta r,
\label{N2}
\end{equation}
that, together with $r'=r + \delta r$, is the infinitesimal version of (\ref{rfpgauges}).

For the physical degree of freedom count we have no Lagrangian constraints ($l=0$), 1 gauge identity ($g=1$) and one effective parameter ($e=1$). Thus, the expression (\ref{29}) gives
\begin{equation}
4-\frac{1}{2}\left(0+1+1\right)=3.
\label{Dan}
\end{equation}
The Hamiltonian counterpart yields $N_1=1$ first-class constraints and $N_2=0$ second-class constraints. This shows the consistency of both procedures.

\subsection{An example of the case (c): the relativistic free particle}\label{sec4.5}
An equivalent Lagrangian action that describes the motion of a relativistic free particle in the Minkowski space-time is given by
\begin{equation}
S[x^\mu, \lambda]=\displaystyle{\int\left[\frac{1}{4\lambda}\dot{x}^\mu\dot{x}^\nu\eta_{\mu\nu}-\lambda m^2c^2\right]dr}.
\label{RP1}
\end{equation}
This action is invariant under reparametrizations of the world line
\begin{eqnarray}\label{simetria}
r' = f(r), \quad {x'}^{\mu} (r') = x^{\mu} (r), \quad \lambda' (r') = \frac{d r}{dr'} \lambda (r).
\end{eqnarray}
The variational derivatives are
\begin{eqnarray}
E_{\alpha}^0 &:=& \frac{d}{dr}\left(\frac{\partial L}{\partial\dot{x}^\alpha}\right)-\frac{\partial L}{\partial x^\alpha}=\frac{1}{2\lambda}\ddot{x}_\alpha-\frac{\dot{\lambda}}{2\lambda^2}\dot{x}_\alpha, \nonumber\\
E_{\lambda}^0 &:=& \frac{d}{dr}\left(\frac{\partial L}{\partial\dot{\lambda}}\right)-\frac{\partial L}{\partial\lambda}=\frac{1}{4\lambda^2}\dot{x}^\alpha\dot{x}_\alpha+m^2c^2.
\label{RP2}
\end{eqnarray}
Therefore
\begin{equation}
E^0:=\left(
\begin{array}{c}
E_\alpha^0\\
E_\lambda^0
\end{array}
\right)=W^0\left(
\begin{array}{c}
\ddot{x}_\alpha\\
\ddot{\lambda}
\end{array}
\right)+K^0,
\label{RP3}
\end{equation}
with
\begin{equation}
W^0=\left(
\begin{array}{c c}
\frac{1}{2\lambda} I_{4X4} & 0\\
0  &  0
\end{array}
\right),\,\,\,\,
K^0=\left(
\begin{array}{c}
-\frac{\dot{\lambda}}{2\lambda^2}\dot{x}_\alpha\\
\frac{1}{4\lambda^2}\dot{x}^\alpha\dot{x}_\alpha+m^2c^2
\end{array}
\right).
\label{RP4}
\end{equation}
This means that we are in the case $(c)$. The Hessian matrix has just one left null vector
\begin{eqnarray}
\Lambda_1 = \left ( 0,0,0,0,1 \right ). \label{RP5}
\end{eqnarray}
The contraction of $\Lambda_1$ with (\ref{RP3}) gives
\begin{equation}
E_\lambda^0 = K^0_{\lambda}.
\label{RP6}
\end{equation}
This is not a gauge identity. From (\ref{RP6}) and $E^0=0$, we get the constraint
\begin{eqnarray}\label{faltaprel}
\psi_0 := K^0_{\lambda} =0.
\end{eqnarray}
End of step 0, which includes $g_0=0$ gauge identities and $l_0=1$ Lagrangian constraint (\ref{faltaprel}).

Step 1. Adding ${\dot \psi}_0=0$ to $E^0=0$ we get
\begin{eqnarray}
E^1:=\left(
\begin{array}{c}
E^0\\
\frac{d}{dr}\psi_0
\end{array}
\right) & =\left(
\begin{array}{c}
E^0\\
\displaystyle{\frac{1}{2\lambda^2}\dot{x}^\alpha\ddot{x}_\alpha-\frac{\dot{\lambda}}{2\lambda^3}\dot{x}^\alpha\dot{x}_\alpha}
\end{array}
\right)\\
       &=
       W^1\left(
\begin{array}{c}
\ddot{x}_\alpha\\
\ddot{\lambda}
\end{array}
\right)+K^1,
\label{RP7}
\end{eqnarray}
with
\begin{equation}
W^1=\left(
\begin{array}{c c}
\frac{1}{2\lambda} I_{4X4} & 0\\
0  &  0\\
\frac{\dot{x}^\alpha}{2\lambda^2} & 0
\end{array}
\right),\,\,\,\,
K^0=\left(
\begin{array}{c}
-\frac{\dot{\lambda}}{2\lambda^2}\dot{x}_\alpha\\
\frac{1}{4\lambda^2}\dot{x}^\alpha\dot{x}_\alpha+m^2c^2\\
-\frac{\dot{\lambda}}{2\lambda^3}\dot{x}^\alpha\dot{x}_\alpha
\end{array}
\right).
\label{RP8}
\end{equation}
The matrix $W^1$ has two left null vectors
\begin{eqnarray}
\lambda_2 = \left ( 0,0,0,0,1,0 \right ), \quad \lambda_3 = \left ( - {\dot x}^{\alpha} / \lambda , 0,1 \right ). \label{RP9}
\end{eqnarray}
$\lambda_2$ corresponds to $\Lambda_1$ augmented by one zero and its use gives again (\ref{RP6}). On the other hand, the contraction of $\lambda_3$ with $E^1$ leads to
\begin{equation}
G^1:=-\frac{\dot{x}^\alpha}{\lambda}E_\alpha^0+\frac{d}{dr} \psi_0 =0.
\label{RP10}
\end{equation}
Using (\ref{RP6}) and (\ref{faltaprel}) we rewrite $\psi_0$ as $\psi_0 = E^0_{\lambda}$ and therefore (\ref{RP10}) is a gauge identity
\begin{equation}
G^1:=-\frac{\dot{x}^\alpha}{\lambda}E_\alpha^0+\frac{d}{dr} E^0_{\lambda} =0.
\label{RP10++}
\end{equation}
End of step 1, which involves $g_1=1$ gauge identities and $l_1=0$ Lagrangian constraints.

Multiplying (\ref{RP10++}) by an arbitrary function $\varepsilon(r)$ and rewriting we get the Noether's identity
\begin{equation}
- \frac{ \varepsilon \dot{x}^\alpha}{\lambda}E_\alpha^0-\dot{\varepsilon}E_\lambda^0+\frac{d}{dr}\left[\varepsilon E_\lambda^0\right]=0,
\label{RP11}
\end{equation}
from which it is possible to read the gauge transformation
\begin{equation}
\delta_{\mbox{gauge}} x^\alpha=- \frac{\varepsilon \dot{x}^\alpha}{\lambda},\hspace{8mm}  \delta_{\mbox{gauge}} \lambda=- \dot{\varepsilon}.
\label{RP12}
\end{equation}
Without loss of generality we can take $\varepsilon := \lambda \delta r$ , and then the gauge transformation looks like
\begin{eqnarray}
\delta_{\mbox{gauge}} x^{\alpha} = - {\dot x}^{\alpha} \delta r, \quad \delta_{\mbox{gauge}} \lambda = - \frac{d}{dr} \left ( \lambda \delta r\right ),
\end{eqnarray}
which, together with $r'=r + \delta r$, is the infinitesimal version of (\ref{simetria}). Notice that the gauge transformation involves $\delta r$ and its time derivative.

In summary, we have $N=5$ configuration variables $(x^\alpha,\lambda)$, $l=l_0 + l_1 = 1+0 =1$ Lagrangian constraint (\ref{faltaprel}), $g=g_0 + g_1 = 0+1=1$ gauge identity (\ref{RP10++}) and $e=2$ effective parameters (\ref{RP12}). Inserting these data into (\ref{29}), we get
\begin{equation}
5-\frac{1}{2}(1+1+2)=3,
\label{RP13}
\end{equation}
physical degrees of freedom, as it must be.

Moreover, the relation (\ref{key2}) and (\ref{key3}) allows us to know the number of first-class constraints $N_1=e=2$ and second-class constraints $N_2=l+g-e=0$ that should appear if the Hamiltonian analysis were done. This is in agreement  with what we get from Dirac's analysis.

\section{\label{sec5}The approach from the covariant canonical formalism viewpoint}
In Section \ref{sec3} we developed a procedure to obtain all the relevant Lagrangian information (Lagrangian constraints, gauge transformations, effective parameters) to count the number of physical degrees of freedom for a Lagrangian system. However, the geometric structure underlying the approach is missing. 
In this section we show that one can understand what is happening geometrically: the geometrical meaning of the constraints and the geometrical significance of the Lagrangian parameters included in the Lagrangian formula for the count of the physical degrees of freedom. Furthermore, it is shown that in order to make the counting it is not necessary to know the gauge transformation of the theory because the presymplectic structure after being restricted to the Lagrangian constraints contains the necessary information to do it.

There exists a geometric approach for revealing the Lagrangian constraints that was first developed in a more general form in the context of global infinite-dimensional symplectic geometry and from the Hamiltonian side, showing that it improves and generalizes the Dirac-Bergman analysis.\cite{Hinds} When we are working in the Lagrangian side  we have to deal with two separate things: the constraint algorithm \cite{Gotay} and the second-order equation problem.\cite{GotayNester} This issues would give rise to the Lagrangian constraints.

Furthermore, in the geometric approach one can only obtain the gauge transformations {\it on-shell} (as degenerate directions of the presymplectic structure),\cite{Wald} which contrasts with the preceding discussion where the gauge transformations are considered {\it off-shell}.

Here, we give a brief summary of the geometric approach based on Refs.~\onlinecite{Gotay,GotayNester} and for the sake of simplicity we work in a natural bundle chart (see Ref.~\onlinecite{Gotay} for a coordinate free formulation).

We take $\cal{C}$ to be the configuration space of some physical system; $T\cal{C}$ is the velocity phase space (tangent bundle). With the help of the Lagrangian function it is possible to define a preferred presymplectic two-form on $T\cal{C}$ written in local coordinates $q^i$ and ${\dot q}^i$ as
\begin{equation}
\Omega = \frac{ \partial^2 L}{ \partial \dot{q}^i \partial q^j }
   dq^i \wedge dq^j +
\frac{ \partial^2 L}{ \partial \dot{q}^i \partial \dot{q}^j }
    dq^i \wedge d\dot{q}^j .\label{gp1}
\end{equation}
(see Ref.~\onlinecite{Wald} for fields). The Lagrangian $L$ is said to be regular iff $\Omega$ is non-degenerate, otherwise $L$ is singular or irregular. Note that $\Omega$ is non-degenerate iff the Hessian $ W^0=\left( \frac{ \partial^2 L}{ \partial \dot{q}^i \partial \dot{q}^j }  \right)$ is invertible. On the other hand, in the singular case, the Legendre map ($FL$) from the tangent bundle $T\cal{C}$ to the cotangent bundle (or phase space ) $T^*\cal{C}$ is no longer invertible. Therefore, there are functions on $T\cal{C}$ that can not be projected to
functions on the phase space.

(i) {\it Algorithm to get the Lagrangian constraints (constraint algorithm).}

When $\Omega$ is non-degenerate the Lagrangian equations of motion (\ref{falta}) can be written in the form
\begin{equation}
X \contr\,\, \Omega - dE= 0, \label{gp2+}
\end{equation}
where $E(q^i, \dot{q}^i )= \dot{q}^i \frac{ \partial L}{ \partial \dot{q}^i  } - L$ is called the ``energy'' (even though it does not need to correspond with a notion of physical energy) and $X$ is a unique vector on the tangent bundle. From this viewpoint, Lagrange's equations (\ref{falta}) are those that correspond to the integral curves of the vector field $X$. On the other hand, when $\Omega$ is degenerate we can still try to write the Lagrange equations of motion (\ref{falta}) as the integral curves of a (to be determined) vector  field $X=\alpha^i \frac{ \partial}{ \partial q^i}+ \beta ^i \frac{ \partial}{ \partial \dot{q}^i}$ on the tangent bundle but we have to be careful because $X$ and $X + Z$ where $Z$ is an arbitrary null vector of $\Omega$ also satisfy
\begin{equation}
X \contr\,\,  \Omega - dE= 0, \label{gp2}
\end{equation}
and so $X$ is not unique. Moreover, notice that there are points on the tangent bundle where $ X \contr\,\, d E \neq 0$. This fact is inconsistent with (\ref{gp2}). The inconsistency is solved by using the constraint algorithm described below, it tells us whether or not these equations have solutions.

This algorithm generates a sequence of sub-manifolds
\begin{equation}
T{\cal{C}} =: P_1 \supseteq P_2 \supseteq P_3 \supseteq \cdots,
\end{equation}
defined by
\begin{eqnarray}
 P_{r+1}:= \{ m \in P_r  \mid  {\phi_r}\textrm{'s} :=  TP_r^\perp \contr \,\, d E  (m)=0 \},
\end{eqnarray}
where
\begin{eqnarray}
TP_r^\perp := \{Z \in T(T{\cal{C}})  \mid  \Omega ( Z ,  TP_r)=0 \},
\end{eqnarray}
in  shorthand notation. The algorithm must end with some final constraint sub-manifold $P := P_s \not = \emptyset $, $1\leq s < \infty$. Thus, on $P$ we have completely consistent equations of motion
\begin{equation}
(X \contr \,\, \Omega - dE) \mid_{P}= 0 \label{gp3}
\end{equation}
and  at least one solution $X \in TP$ exists. Note that the solutions to (129) may not be unique,  we can add to the solution a vector field in $\ker \Omega\cap TP$ and still it will be a solution. The final constraint submanifold $P$ is maximal.

For practical calculations note that $TP_1^\perp= \ker \Omega$, if $Z\in TP_1^\perp$, then $Z \contr \,\, X \contr \,\, \Omega= \Omega(X,Z) =0$, then (\ref{gp2}) require that $Z \contr \,\, dE =0$. Therefore, the points of $T\cal{C}$ where Eqs. (\ref{gp2}) are inconsistent are those for which $Z \contr \,\, dE \not=0$ for any $Z \in \ker \Omega$. Thus
\begin{equation}
P_2= \{m \in  T{\cal{C}} \mid  {\phi_1} 's := \ker  \Omega \contr \,\,  d E\  (m)=0 \}.
\end{equation}

We now try to solve
\begin{equation}
(X \contr \,\, \Omega - dE)\mid_{P_2}= 0.
\end{equation}
This equation can be solved algebraically for $X$, but also, physically, we must demand that the motion of the system takes place on $P_2$, $X$ must be tangent to $P_2$; this requirement is not automatically accomplished, generating more Lagrangian constraints.
That is the origin of $P_3, P_4$, etc. For practical calculations it is much better to use that $X$ is tangent to $P_2$ iff $X({\phi_1} 's )=0$, this requirement gives rise to some constraints ${\phi_2} 's$ which define $P_3$, then we require $X$ to be tangent to $P_3$, and so on.

In this way we get $N_1+N_2-N^{(p)}$ Lagrangian constraints (where $N^{(p)}$ is the number of primary constraints) that correspond to the projectable ones; this means that $FL (\phi 's)$ is a Hamiltonian constraint (all the secondary ones).\cite{Gotay}

(ii) {\it The second-order equation problem.}

Variational as well as physical considerations require that the Lagrange equations (\ref{gp2}) be a set of {\it second-order} differential equations.\cite{Nester} This requirement means
\begin{equation}\label{bogard}
X (q^i)= \dot{q}^i \Longleftrightarrow  \alpha^i=\dot{q}^i.
\end{equation}
This condition  together with (\ref{gp2}) generate more Lagrangian constraints ($\phi 's$) and they must satisfy $X({\phi 's} )=0$, and so on. The new Lagrangian constraints are the strictly non-projectable ($N^{(p)}-N^{(p)}_1$).\cite{Pons} In the regular case (\ref{bogard}) is not imposed because (\ref{gp2+}) always implies $\alpha^i=\dot{q}^i$.\cite{GotayNester}

Therefore it is possible to obtain all the Lagrangian  constraints $l  = (N_1+N_2-N^{(p)} )+ (N^{(p)}-N^{(p)}_1) = N_1 + N_2 - N^{(p)}_1$, and their origin is clear (if they come from the constraint algorithm or the second-order problem), one advantage is that we know which constraints are the projectable ones and which are not.

We could now compute the number of degrees of freedom, this corresponds to  $\frac{1}{2} \mbox{Rank} \,\,\Omega \mid_{\phi 's}$ (Ref.~\onlinecite{Teitelboim} for the Hamiltonian side, chapter 2), something equivalent was shown in Ref.~\onlinecite{Gracia}: it is shown that after taking into account the Lagrangian constraints one needs $N_1+N^{(p)}_1$ conditions to fix the norm on the Lagrangian side, that means that $\Omega \mid_{\phi \textrm{'s}}$  has   $N_1+N^{(p)}_1$ null vectors, therefore $\frac{1}{2}  \mbox{Rank} \,\,  \Omega \mid_{\phi 's}= \frac{1}{2}(2N-l-(N_1+N^{(p)}_1))$ and using the results 1 and 2 of Section \ref{sec3} ($N_1=e$, $  N^{(p)}_1=g$) $\frac{1}{2}  \mbox{Rank} \,\, \Omega \mid_{\phi 's}=  N-\frac{1}{2}\left(l+g+e\right)$, which of course coincides with (\ref{29}). Thus the geometrical meaning
of  $g+e$ is the number of null vectors of $\Omega \mid_{\phi 's}$.

Obtaining the Lagrangian constraints is not the whole story, we should be able to get the gauge transformations in order to identify the parameters $g$ and $e$ (note, if we know the rank of the restricted presymplectic structure then we know $g+e$). This is possible (at least {\it on-shell}), in fact is well known that the gauge transformations
are degenerate directions of the presymplectic form (\ref{gp1}) over the space of solutions,\cite{Wald,Mauricio} equivalently, one can look for the degenerate directions
\begin{equation}
(\tilde{X} \contr \,\, \Omega) \mid_ {\phi 's}=0, \,\mbox{with} \, \tilde{X}=(\delta_\varepsilon q^i) \frac{\partial}{\partial q^i}+ (\delta_\varepsilon \dot{q}^i) \frac{\partial}{\partial \dot{q}^i},  \label{gp4}
 \end{equation}
and  $\delta_\varepsilon q^i, \delta_\varepsilon \dot{q}^i$ must satisfy the $l$ Lagrangian constraints
($\delta_\varepsilon \phi 's = 0$). We now illustrate the procedure using the examples of Section \ref{sec4} (see also Appendix C).

\subsection{Example of the SubSection \ref{sec4.1}}
Consider the Lagrangian system given in (\ref{30})
\begin{equation*}
L(q^i,\dot{q}^i)=\dot{q}^1 q^2 - \dot{q}^2 q^1- (q^1 - q^ 2) q^3,
\end{equation*}
from which
\begin{eqnarray}
\Omega &= &  2 dq^1 \wedge dq^2,\nonumber \\
 E&=& q^3(q^1-q^2),
 \end{eqnarray}
and so $dE= (q^1 - q^2) dq^3 + q^3 (dq^1- d q^2)$. A basis of $\ker \Omega$ is given by
$\displaystyle \left\{ \frac{\partial}{\partial q^3},\frac{\partial}{\partial \dot{q}^i }  \right\}$, $i=1, 2,3$. Nevertheless, notice that among the elements of this set only $\displaystyle Z:=\frac{\partial}{\partial q^3}$ generates a Lagrangian constraint given by $\phi_1=Z \contr \, dE= q^1-q^2 =0$. Continuing with the approach, we must demand that $X (\phi_1)=0$, but because $X$ satisfies (\ref{gp2}) (see (\ref{gp5}) below) this is automatically fulfilled. Thus, the constraint algorithm gives us just one Lagrangian constraint, which is projectable.

We now look for non-projectable Lagrangian constraints. Following the procedure, Eqs. (\ref{gp2}) become:
\begin{eqnarray}
2\alpha^1 +q^3& = & 0,  \nonumber\\
-2\alpha^2 -q^3& = & 0, \nonumber\\
-(q^1-q^2) &=&0. \label{gp5}
\end{eqnarray}
and the requirement $\alpha^i= {\dot q}^i$ implies that $\dot{q}^i$ must satisfy the new Lagrangian constraints
\begin{eqnarray}
\phi_2&:=&2\dot{q}^1 +q^3 =  0, \nonumber\\
\phi_3&:=&-2\dot{q}^2 -q^3 =  0. \label{gp6}
\end{eqnarray}
These are non-projectable. That $X$ overrides these constraints is a consequence of the Lagrange's equations of motion ($\beta^i= \ddot{q}^i$) and the fact the motion takes place on the constraint surface ${\dot \phi}_2=0$ and ${\dot \phi}_3=0$. More precisely,  $X(\phi_2)=\dot{\phi_2}=0, X(\phi_3)=\dot{\phi_3}=0$. Summarizing, we have $l=3$
Lagrangian constraints (see SubSection \ref{sec4.1}), and if we restrict $\Omega$ to them, we have
\begin{eqnarray}
\Omega \mid_{\phi 's}=0,
\end{eqnarray}
and therefore the number of physical degrees of freedom is
\begin{eqnarray}
\frac{1}{2} \mbox{Rank}\,\, \Omega \mid_{\phi 's}=0.
\end{eqnarray}

In order to get the gauge transformation, we have to compute (\ref{gp4}),
\begin{eqnarray}
0=( {\tilde X} \contr \,\, \Omega) \mid_{\phi 's}= 2(\delta_\varepsilon q^1-\delta_\varepsilon q^2)dq^1.
\end{eqnarray}
Therefore $\delta_\varepsilon q^1=\delta_{\varepsilon} q^2$,
and taking into account (\ref{gp6}), $\delta_\varepsilon q^3= -2 \delta_\varepsilon \dot{q}^1$,
hence by redefining $\delta_\varepsilon q^2= \frac{1}{2}\varepsilon(t)$ the gauge transformation reads
\begin{equation*}
\delta_\varepsilon q^1=\frac{\varepsilon}{2},
\hspace{1cm}
\delta_\varepsilon q^2=\frac{\varepsilon}{2},
\hspace{1cm}
\delta_\varepsilon q^3=-\dot{\varepsilon},
\end{equation*}
in agreement with (\ref{37}).

\subsection{Example of the SubSection \ref{sec4.2}}
Let us consider the Lagrangian system (\ref{LH1})
\begin{equation*}
L(q^i,\dot{q}^i)=(\dot{q}^1+\dot{q}^2) q^3 + \frac{1}{2} \left ( \dot{q}^3 \right)^2.
\end{equation*}
Thus
\begin{eqnarray}
\Omega &= & (dq^1+ dq^2) \wedge dq^3 +dq^3 \wedge d\dot{q}^3, \nonumber\\
 E&=& \frac{1}{2} \left ( \dot{q}^3 \right)^2 \Rightarrow dE= \dot{q}^3 d \dot{q}^3.
\end{eqnarray}
A basis of  $\ker \Omega$ is $\{ Z_1, Z_2, Z_3, Z_4 \}$ with
\begin{eqnarray}
Z_1 &:=& \frac{\partial}{\partial q^1}+ \frac{\partial}{\partial \dot{q}^3}, \quad
Z_2 := \frac{\partial}{\partial q^2}+ \frac{\partial}{\partial \dot{q}^3},\nonumber\\
Z_3 &:=& \frac{\partial}{\partial \dot{q}^1 }, \quad Z_4:= \frac{\partial}{\partial \dot{q}^2 }.
\end{eqnarray}
Notice that just $ Z_1, Z_2$ give rise to Lagrangian constraints. Nevertheless, they generate the same constraint
$\phi_1= \dot{q}^3=0$. We now demand that $X (\phi_1)=0$ and because $X$ satisfies (\ref{gp2}) (see also (\ref{gp7}) below), we have
\begin{eqnarray}
0=X (\phi_1)=\alpha^1 + \alpha^2=:\phi_2,
\end{eqnarray}
but this expression is unsuitable, because we do not have $\alpha^i$ in terms of $q^i, \dot{q}^i$. Continuing, if we now use $\alpha^i=\dot{q}^i$, then $\phi_2$ becomes $\phi_2=\dot{q}^1+ \dot{q}^2=0$. Because we have used $\alpha^i=\dot{q}^i$ then $\phi_2$ is a non-projectable constraint. That $X(\phi_2)=0$ is a consequence of ${\dot \phi}_2=0$ and the equations of motion, because $X(\phi_2)= \dot{\phi}_2$.

Up to here, we have two Lagrangian constraints, $\phi_1$ which is projectable and $\phi_2$ which is non-projectable.

There could be more non-projectable constraints, Eqs. (\ref{gp2}) become
\begin{eqnarray}
\alpha^3 & = & 0,  \nonumber\\
(\alpha^1 + \alpha^2)-\beta^3 &= & 0, \nonumber\\
\alpha^3 -\dot{q}^3 &=&0.  \label{gp7}
\end{eqnarray}
and the requirement $\alpha^i=\dot{q}^i$ implies that $\dot{q}_i$ must satisfy
\begin{eqnarray}
\dot{q}^3 =  0, \nonumber\\
\beta^3-(\dot{q}^1+ \dot{q}^2) =  0.
\end{eqnarray}
The first one is just $\phi_1$ and the second one, taking into account that $\beta^3=\ddot{q}^3$, reduces to $\phi_2$.

Therefore, we have just one non-projectable Lagrangian constraint.

Thus, we have $l=2$ Lagrangian constraints (in agreement with SubSection \ref{sec4.2}), and restricting $\Omega$ to them gives
\begin{eqnarray}
 \Omega \mid_{\phi 's}=  (dq^1+ dq^2) \wedge  dq^3,
\end{eqnarray}
and so the number of physical degrees of freedom is
\begin{eqnarray}
\frac{1}{2} \mbox{Rank} \,\,\Omega \mid_{\phi 's}=\frac{1}{2} 2=1.
\end{eqnarray}

We now look for the gauge transformation. Equation (\ref{gp4}) acquires the form
\begin{eqnarray}
0&=& ( {\tilde X} \contr \,\, \Omega) \mid_{\phi\textrm{'s}}\nonumber\\
&=& (\delta_\varepsilon q^1+ \delta_\varepsilon q^2)dq^3-\delta_\varepsilon q^3(dq^1+dq^2) .
\end{eqnarray}
By inspection, we get  $\delta_\varepsilon q^1=-\delta_\varepsilon q^2$ and $\delta_\varepsilon q^3=0$, and from these we get  $\delta_\varepsilon \phi_1=0=\delta_\varepsilon \phi_2$; hence by redefining $\delta_\varepsilon q^1= \varepsilon(t)$ the gauge transformation reads
\begin{eqnarray}
\delta_\varepsilon q^1=\varepsilon,
\hspace{1cm}
\delta_\varepsilon q^2=-\varepsilon,
\hspace{1cm}
\delta_\varepsilon q^3=0, \nonumber
\end{eqnarray}
in agreement with the result of SubSection \ref{sec4.2} (see Eq. (\ref{LH9})).

\subsection{Example of the SubSection \ref{sec4.3}}
For the Lagrangian system (\ref{H4}), the presymplectic two-form and the ``energy'' function are
\begin{eqnarray}
\Omega &=& e^{q^1} dq^1\wedge dq^2 +dq^2\wedge dq^3+dq^2\wedge d\dot{q}^2+dq^3\wedge d\dot{q}^3, \nonumber\\
 E      &=& \frac{1}{2}\left[\left(\dot{q}^2\right)^2+\left(\dot{q}^3\right)^2-\left(q^2\right)^2-e^{2q^1}\right].
 \label{H37}
\end{eqnarray}
and so $dE=\dot{q}^2d\dot{q}^2+\dot{q}^3d\dot{q}^3-q^2dq^2-e^{2q^1}dq^1$. A basis of  $\ker \Omega$ is $\{Z_1, Z_2\}$ with
\begin{equation}
Z_1:= e^{-q^1}\frac{\partial}{\partial q^1}+\frac{\partial}{\partial\dot{q}^2}, \hspace{5mm}
Z_2:= \frac{\partial}{\partial\dot{q}^1}.
\label{H38}
\end{equation}
Notice that just $Z_1$ gives rise to the Lagrangian constraint $\phi_1:= {Z_1} \, \contr \,\, dE=\dot{q}^2-e^{q^1}=0$. We now demand $X(\phi_1)=0$, which leads us to the condition
\begin{equation}
X(\phi_1)=\beta^2-\alpha^1 e^{q^1}=0.
\label{H39}
\end{equation}
Now, let us analyze if (\ref{H39}) is automatically satisfied or imposes any restriction. Recalling that $X$ must satisfy the equation of motion (\ref{gp2+}), it follows that
\begin{eqnarray}
e^{q^1}\left(e^{q^1}-\alpha^2\right) & = & 0,  \nonumber\\
\beta^2-e^{q^1}\alpha^1 +\alpha^3-q^2 &= & 0, \nonumber\\
\alpha^2 -\beta^3 &=&0,\nonumber\\
\alpha^2-\dot{q}^2 &=&0,\nonumber\\
\alpha^3-\dot{q}^3 &=&0.
\label{H40}
\end{eqnarray}
The substitution of $\beta^2$ and $\alpha^3$ given in (\ref{H40}) into (\ref{H39}) gives the new constraint
\begin{equation}
\phi_2:=-\dot{q}^3+q^2=0.
\label{H42}
\end{equation}
Up to here we have two projectable Lagrangian constraints $\phi_1$ and $\phi_2$, a fact that can be easily verified by substituting $\phi_1$ and $\phi_2$ into (\ref{H29}), which implies the Hamiltonian constraints $\phi_2$ given in (\ref{H33}) and $\phi_3$ given in (\ref{H34}).\\
Nevertheless, there could be non-projectable constraints as consequence of (\ref{H40}). The requirement $\alpha^i=\dot{q}^i$ into (\ref{H40}) implies that $\dot{q}^i$ must satisfy
\begin{eqnarray}
e^{q^1}\left(e^{q^1}-\dot{q}^2\right) & = & 0,  \nonumber\\
\beta^2-e^{q^1}\dot{q}^1 +\dot{q}^3-q^2 &= & 0, \nonumber\\
\dot{q}^2-\beta^3 &=&0.
\label{H44}
\end{eqnarray}
The first one is just $-\phi_1=0$. Because $\phi_1=0$ implies $\dot{q}^2=e^{q^1}$, then $\beta^2=\ddot{q}^2=\dot{q}^1e^{q^1}$ and the second one reduces to $-\phi_2$. Since $\phi_2=-\dot{q}^3+q^2=0$, the third one ($-\ddot{q}^3+\dot{q}^2=0$) follows from $\dot{\phi}_2=0$. Therefore, we do not have non-projectable Lagrangian constraints.

Thus, we have $l=2$ Lagrangian constraints (in agreement with the Lagrangian analysis), and restricting $\Omega$ to them gives
\begin{eqnarray}
 \Omega \mid_{\phi 's}=0,
 \label{H45}
\end{eqnarray}
and so the number of physical degrees of freedom is
\begin{eqnarray}
\frac{1}{2} \mbox{Rank} \,\,\Omega \mid_{\phi 's}=0.
\label{H46}
\end{eqnarray}
We now look for the gauge transformation. Equation (\ref{gp4}) acquires the form
\begin{eqnarray}
0 &=& ( {\tilde X} \contr \,\, \Omega) \mid_{\phi 's}\nonumber\\
&=& \left(e^{q^1}\delta_\varepsilon q^1-\delta_\varepsilon \dot{q}^2\right)dq^2+\left(\delta_\varepsilon q^2-\delta_\varepsilon \dot{q}^3\right)dq^3.
\label{H47}
\end{eqnarray}
By inspection, we get  $\delta_\varepsilon q^1=e^{-q^1}\delta_\varepsilon \dot{q}^2$ and $\delta_\varepsilon q^2=\delta_\varepsilon \dot{q}^3$,
and from these we get  $\delta_\varepsilon \phi_1=0=\delta_\varepsilon \phi_2$; hence by redefining $\delta_\varepsilon q^3= \varepsilon(t)$ the gauge transformation reads
\begin{eqnarray}
\delta_\varepsilon q^1=e^{-q^1}\ddot{\varepsilon},
\hspace{1cm}
\delta_\varepsilon q^2=\dot{\varepsilon},
\hspace{1cm}
\delta_\varepsilon q^3=\varepsilon,
\label{H48}
\end{eqnarray}
in agreement with the result of the Lagrangian analysis (see Eq. (\ref{H26})).

\subsection{Example of the SubSection \ref{sec4.4}}
The starting point is the action (\ref{rfpa})
\begin{equation*}
S[x^\mu ]=-mc\int_{r_1}^{r_2}\sqrt{-\eta_{\mu\nu}\frac{dx^\mu(r)}{dr}\frac{dx^\nu(r)}{dr}}dr,
\end{equation*}
which leads to the presymplectic two-form $\Omega$ and the energy $E$
\begin{eqnarray}
\Omega_1 &= & A_{\mu \nu} dx^{\mu} \wedge d {\dot x}^{\nu} \nonumber\\
 &=& \left ( \frac{mc \eta_{\mu\nu} \dot{x}^\alpha \dot{x}_\alpha- mc\dot{x}_\mu \dot{x}_\nu } {\sqrt{-\eta_{\alpha\beta} \dot{x}^\alpha \dot{x}^\beta  }} \right ) d x^{\mu} \wedge d {\dot x}^{\nu} , \nonumber\\
 E&=& 0 \Rightarrow dE=0.
\end{eqnarray}
A basis of $\ker \Omega_1$ is
$\displaystyle \left\{ \dot{x}^{\mu} \frac{\partial}{\partial x^{\mu}}, \dot{x}^{\mu} \frac{\partial}{\partial \dot{x}^{\mu} }  \right\}$ and because of $dE=0$ there are no (projectable) Lagrangian constraints.

Continuing with the algorithm, one set of Eqs. (\ref{gp2}) acquires the form
\begin{eqnarray}
A_{\mu \nu} \alpha^{\mu}& = & 0, \label{gp8}
\end{eqnarray}
we now require $\alpha^{\mu}=\dot{x}^{\mu}$ in Eq. (\ref{gp8}), but $A_{\mu \nu} \dot{x}^{\mu}$ identically vanishes, so there are not any non-projectable Lagrangian constraints. Therefore, we have $l=0$ Lagrangian constraints in agreement with the result of SubSection \ref{sec4.4}. Thus $\mbox{Rank}\,\, \Omega_1=6,$ and the number of physical degrees of freedom is
\begin{eqnarray}
\frac{1}{2} \mbox{Rank} \,\, \Omega_1= \frac{6}{2}=3.
\end{eqnarray}

Finally, in order to get the gauge transformation, we compute
\begin{eqnarray}
0=(\tilde{X} \contr \,\, \Omega_1)=A_{\mu \nu}\delta_\varepsilon x^{\mu} d \dot{x}^{\nu}-A_{\mu \nu}\delta_\varepsilon\dot{x}^{\mu} dx^{\nu},
\end{eqnarray}
and from inspection $\delta_\varepsilon x^{\mu}\propto  \dot{x}^{\mu}$ and $ \delta_\varepsilon \dot{x}^{\mu}\propto \dot{x}^{\mu}$, hence by redefining $\delta_{\varepsilon} x^{\mu}=-\dot{x}^{\mu}\delta r $ (with $\varepsilon=\delta r$) in agreement with the result of SubSection \ref{sec4.4} (see Eq. (\ref{N2})). Of course, it also satisfies $\delta_\varepsilon \dot{x}^{\mu}\propto \dot{x}^{\mu}$ (on-shell).

\subsection{Example of the SubSection \ref{sec4.5}}
Consider now the action principle (\ref{RP1})
\begin{equation*}
S[x^\mu, \lambda]=\displaystyle{\int\left[\frac{1}{4\lambda}\dot{x}^\mu\dot{x}^\nu\eta_{\mu\nu}-\lambda m^2c^2\right]dr}.
\end{equation*}
It leads to
\begin{eqnarray}
\Omega_2 &= & -\frac{1}{2\lambda^2}\eta_{\mu \nu}  \dot{x}^{\mu}  dx^{\nu} \wedge d \lambda+ \frac{\eta_{\mu \nu}}{2 \lambda}dx^{\mu} \wedge d \dot{x}^{\nu} , \nonumber\\
 E&=& \frac{\eta_{\mu \nu} \dot{x}^{\mu} \dot{x}^{\nu}}{4\lambda} + \lambda m^2c^2, \nonumber \\
 \Rightarrow dE&=& \left( m^2c^2-\frac{\eta_{\mu \nu} \dot{x}^{\mu} \dot{x}^{\nu}}{4\lambda^2}\right) d\lambda+
 \frac{\eta_{\mu \nu} \dot{x}^{\mu} }{2\lambda} d \dot{x}^{\nu}.
\end{eqnarray}
A basis of $\ker \Omega_2$ is
$\displaystyle \left\{  \frac{\partial}{\partial \lambda}+ \frac{\dot{x}^{\mu}}{\lambda} \frac{\partial}{\partial \dot{x}^{\mu} }
 ,  \frac{\partial}{\partial \dot{\lambda}}  \right\}$ but only the first one generates a Lagrangian constraint
 \begin{eqnarray}
 \phi=\frac{\eta_{\mu \nu} \dot{x}^{\mu} \dot{x}^{\nu}}{4\lambda^2} + m^2c^2=0.
 \end{eqnarray}
That $X(\phi)=0$ is a consequence of $\dot{\phi} = 0$ and the equations of motion, because $X(\phi)= \dot{\phi}$. Thus, the constraint algorithm gives us just one Lagrangian constraint, which is projectable.

Finally, if we require $\alpha^{\mu}=\dot{x}^{\mu}, \alpha=\dot{\lambda}$ in (\ref{gp2}) it identically vanishes; then no more constraints are generated. Thus, there are not any non-projectable Lagrangian constraints.

Therefore, we have $l=1$ Lagrangian constraints (in agreement with SubSection \ref{sec4.5}). We have $\Omega_2|_{\phi}= \Omega_1 $, then 
$\mbox{Rank}\,\, \Omega_2|_{\phi}=6,$ and the number of physical degrees of freedom is
$$\frac{1}{2} \mbox{Rank} \,\, \Omega_2|_{\phi}= \frac{6}{2}=3 .$$

We now look for the gauge transformation. Equation
(\ref{gp4}) acquires the form
\begin{eqnarray}
0&=&(\tilde{X} \contr \,\, \Omega_2)|_{\phi} \nonumber \\
 &=&
\left( - \frac{\dot{x}_{\mu}\dot{x}_{\nu}}{2\lambda \dot{x}^{\alpha} \dot{x}^{\alpha}}
+\frac{\eta_{\mu}\nu}{2\lambda}\right)  \delta_\varepsilon x^{\mu} d x^{\nu}  \nonumber \\
&& + \left(  \frac{\eta_{\mu}\nu}{2 \lambda} \tilde{\delta} \dot{x}^{\mu}
-\frac{ \eta_{\mu}\nu \dot{x}^{\mu} }{2 \lambda^2}\delta_\varepsilon \lambda
\right)  dx^{\nu}.
\end{eqnarray}
Therefore $\delta_\varepsilon x^{\mu}\propto  \dot{x}^{\mu}, \delta_\varepsilon\lambda= \frac{\lambda}{\dot{x}^{\mu}}\delta_\varepsilon\dot{x}^{\mu}$, hence by redefining $\delta_\varepsilon x^{\mu}=-\dot{x}^{\mu}\delta r $ (with $\varepsilon=\delta r$), then
$\delta_\varepsilon \lambda = - \frac{d}{dr} \left ( \lambda \delta r \right )$, where we have made use of $\phi$,
in agreement with the result of SubSection \ref{sec4.5}.

Even though in the previous examples we were able to obtain the ``same'' gauge transformations than those of the Section \ref{sec4} (in this section $\delta_\varepsilon q^i$, $\delta_\varepsilon \dot{q}^i$ are solutions of the constraints), this is not always possible, this is because  the gauge transformations ({\it on-shell}) not necessarily coincide with the gauge transformation ({\it off-shell}), see Appendix \ref{appC}. Another possible discrepancy is attached to the choice of the vectors of the different basis employed in the two approaches.


\section{\label{sec6}Concluding remarks}
In this paper we have reported a method involving a formula to count the number of physical degrees of freedom of a Lagrangian system employing only Lagrangian parameters.  The approach has the advantage that there is no need to go into Dirac's canonical formalism to make the counting. Other advantage is that in spite of being a Lagrangian method, there is no need to go into all the details of the geometry involved in the covariant canonical formalism. Indeed, the method is robust enough to give information about the number of first-class and second-class constraints without having to perform neither Dirac's canonical analysis nor the covariant canonical formalism.

Finally, further investigation is needed to see how the approach works with fermions and to compare it with the case of bosonic variables developed in this paper. Furthermore, it would be also interesting to apply the approach to field theory and theories of gravity.  Nevertheless, we expect minor and natural modifications that are always involved in these cases. 

\begin{acknowledgments}
Warm thanks to J.D. Vergara for his valuable comments on the subject of this paper. This work was supported in part by CONACyT, M\'exico, Grant Numbers 167477-F and 132061-F.
\end{acknowledgments}

\appendix
\section{\label{appA}Algorithm for getting the gauge symmetries, further details}
Section \ref{sec2} makes use of a general structure for the gauge identities (\ref{GI}) at step k-th in the consistency algorithm. Nevertheless there is no proof of such structure there. This appendix fills out this gap and proves (\ref{GI}) providing, by the way, a general scheme to work out gauge identities.\\
Let us start by recalling that in the k-th step of the procedure described in Section \ref{sec2} we have the following situation:
\begin{equation}
E_{i_k}^{k}:=\left\{
\begin{array}{c l}
W_{ij}^{0}\ddot{q}^j+K_i^{0}, &\hspace{5mm} i=1,\ldots ,N,\\
\frac{d}{dt}\psi_{\bar{a}_1}(q,\dot{q})\,, &\hspace{5mm} \bar{a}_1=1,\ldots ,R_1-g_0,\\
\vdots & \\
\frac{d}{dt}\psi_{\bar{a}_k}(q,\dot{q})\,, &\hspace{5mm} \bar{a}_k=1,\ldots ,R_k-g_{k-1},
\end{array}
\right.
\label{17}
\end{equation}
where $i_k=1,\ldots ,N+(R_1-g_0)+\ldots+(R_k-g_{k-1})$; $\psi_{\bar{a}_1}(q,\dot{q})=0$ are independent Lagrangian constraints that emerge at step 0, $\psi_{\bar{a}_2}(q,\dot{q})=0$ are independent Lagrangian constraints that emerge at step 1, and so on.\\
Once again (\ref{17}) can be summarized as
\begin{equation}
E_{i_k}^{k}:= W_{i_kj}^{k}\ddot{q}^j+K_{i_k}^{k}.
\label{18}
\end{equation}
Now, we look for left null vectors of $W^{k}$. These vectors include those of the previous step augmented by an appropriate number of zeros and their contraction with $E^k$ gives relations that we already have in steps behind. Taking out these ones, let us suppose there are $R_{k+1}$ and called $\lambda_{a_{k+1}}$. Their contractions with $E^{k}$ produce the functions
\begin{equation}
\lambda_{a_{k+1}}^{i_k}E_{i_k}^{k}=\lambda_{a_{k+1}}^{i_k}(q,\dot{q})K_{i_k}^{k}(q,\dot{q}),
\hspace{0.5cm}
a_{k+1}=1,\ldots ,R_{k+1},
\label{13}
\end{equation}
which vanish imposing $E^k=0$ and $\psi_{\bar{a}_m}=0$ for $1\leq m\leq k$(``on-shell"). Nevertheless, it is possible that not all functions in (\ref{13}) are independent among themselves nor with the set of independent functions of the steps above, $\psi_{\bar{a}_1},\ldots,\psi_{\bar{a}_k}$. This lead us to ${R}_{k+1}-g_k$ \emph{new} independent functions $\psi_{\bar{a}_{k+1}}(q,\dot{q})$ and, by mean of general dependence, $g_k$ new independent nontrivial relations
\begin{equation}
C^{a_{k+1}}_{\mathfrak{g}_k}\lambda_{a_{k+1}}^{i_k}E_{i_k}^{k}=C^{\bar{a}_m}_{\mathfrak{g}_k}\psi_{\bar{a}_m},\,\,\left\{
\begin{array}{l}
\mathfrak{g}_k=1,\ldots,g_k,\\
m=1,\ldots,k,
\end{array}
\right.
\label{GAT1}
\end{equation}
for appropriate coefficients $C$ that, in general, depend on $q$'s and time derivatives thereof. The rearrangement of the relations in (\ref{GAT1}) leads to the ``off-shell" relation
\begin{equation}
G_{\mathfrak{g}_k}^{k}:=C^{a_{k+1}}_{\mathfrak{g}_k}\lambda_{a_{k+1}}^{i_k}E_{i_k}^{k}-C^{\bar{a}_m}_{\mathfrak{g}_k}\psi_{\bar{a}_m}=0,
\label{GAT2}
\end{equation}
which are gauge identities.

Now, taking into account (\ref{17}), expression (\ref{GAT2}) can be rewritten as
\begin{equation}
G_{\mathfrak{g}_k}^{k}=C^{a_{k+1}}_{\mathfrak{g}_k}\left[\lambda_{a_{k+1}}^{i}E_{i}^{0}+\lambda_{a_{k+1}}^{\bar{a}_m}\frac{d}{dt}\psi_{\bar{a}_m}\right]-C^{\bar{a}_m}_{\mathfrak{g}_k}\psi_{\bar{a}_m}=0.
\label{GAT3}
\end{equation}
Now we go back step by step and recover the form of $\psi_{\bar{a}_m}$ in terms of $E^0$ and their time derivatives(see examples). This is always possible since each $\psi_{\bar{a}_1}$ comes from contractions in the form (\ref{falta2}), which can be thought as relations between the variational derivatives $E^0_i$. $\psi_{\bar{a}_2}$ comes from combinations of contractions similar to (\ref{falta2}):
\begin{equation}
\lambda_{a_2}^{i_1}E^1_{a_1}=\lambda_{a_2}^{i_1}K^1_{a_1}.
\label{Nat1}
\end{equation}
Moreover, $(E^1)^T=\left(E^0,\frac{d}{dt}\psi_{\bar{a}_1}\right)$ and then (\ref{Nat1}) becomes
\begin{equation}
\lambda_{a_2}^iE^0_i+\lambda_{a_2}^{\bar{a}_1}\frac{d}{dt}\psi_{\bar{a}_1}=\lambda_{a_2}^{i_1}K^1_{a_1}.
\label{Nat2}
\end{equation}
Thus, such combination $\psi_{\bar{a}_2}$ can be expressed in terms of $E^0_i$ and $\frac{d}{dt}\psi_{\bar{a}_1}$, which could be thought entirely in terms of $E^0_i$ and their first time derivatives by rewriting $\psi_{\bar{a}_1}$ in terms of $E^0_i$ as we described before. Something similar occurs for $\psi_{\bar{a}_3}$, with the corresponding increase in the order of the time derivative, and so on.\\
Substitution of $\psi_{\bar{a}_m}$ into (\ref{GAT3}), and appropriate use of the product rule for derivatives to construct total derivatives leads to
\begin{equation*}
G_{\mathfrak{g}_k}^{k}=\sum_{s=0}^{k}\frac{d^s}{dt^s}\left({M^{(k)}}_s^iE_i^{0}\right)=0,
\end{equation*}
where ${M^{(k)}}_s^i$ are some functions of $q$'s and their derivatives, emerging from $C_{\mathfrak{g}_k}$ and $\lambda_{a_{k+1}}$ given by the specific theory.

This proves (\ref{GI}) and the rest of the procedure continues as it was described in Section \ref{sec2}.

\section{\label{appB}Proof of the results 1 and 2 of Section \ref{sec3}}
It follows the proof of the results 1 and 2 of Section \ref{sec3}. Let us denote by $\Omega$ the Hamiltonian constraints. In order to simplify our notation we will add an index that describes the level of appearance of each constraint in the consistency algorithm. Thus, for instance, any secondary constraint $\Omega_p$ that appears at level $s$ in the consistency algorithm will be denoted $\Omega_{p_s}$ and the primary constraints $\Omega_\alpha$ will be noted $\Omega_{\alpha_0}$. The convention sum is adopted in the rest of the paper and the sum symbol will be used when considered necessary to clarify arguments.

Now, we will find the local gauge symmetries of $S_T$. In terms of \emph{Dirac's} theory,\cite{Teitelboim} the dynamics of the total formalism is ruled by $H_T=H^{(1)}+v^{\beta_0}_{(1)}\Omega_{\beta_0}^{(1)}$, where the superscript $(1)$ means first-class. The consistency requirement for constraints tells us
\begin{equation}
\dot{\Omega}_{\alpha_0}=\{\Omega_{\alpha_0} ,H^{(1)}\}+v^{\beta_0}_{(1)}\{\Omega_{\alpha_0},\Omega_{\beta_0}^{(1)}\}\approx 0.
\label{21}
\end{equation}
It is noteworthy that the consistency algorithm is subtle and all kind of possibilities can arise in principle (first-class constraints generate second-class constraints or vice versa, there is or not a global separation in first- and second-class constraints, etc).\cite{Teitelboim} However many of this complications are atypical and we can establish some condition that are satisfied in most of the cases and illustrate the general behavior of the theory. We will assume that:

(i) The rank of the Poisson bracket matrix of constraints is constant on the constraint surface $\Gamma$, i.e., $\mbox{Rank} \left[\{\Omega_{A},\Omega_{B}\}\right]$ constant on $\Gamma$, where $A$ and $B$ run in the whole set of constraints. This guarantees a global separation in first- and second-class constraints and that the generation to which a constraint belongs is well defined.

(ii) The first- and second-class constraints are not mixed in the consistency algorithm. This means that the Poisson brackets of first-class constraints do not involve squares of second-class constraints and the consistency conditions $\dot{\Omega}_{p_i}=0$ for second class constraints do not generate first-class constraints.

(iii) The first-class constraints are irreducible. This is done in order to have the theory simpler and clear as possible.

As result of the assumption (i) and the fact that $\Omega_{\beta_0}^{(1)}$ are first-class quantities, on $\Gamma_1$
\begin{equation}
\{\Omega_{\alpha_0},\Omega_{\beta_0}^{(1)}\}=\bar{C}_{\alpha_0\beta_0}^{\gamma_0}\Omega_{\gamma_0}.
\end{equation}
Using again (i) and since $\{\Omega_{\alpha_0},H^{(1)}\}\approx \bar{V}_{\alpha_0}^{p_1}\Omega_{p_1}$, the constructive methodology guarantees that
\begin{equation}
\{\Omega_{\alpha_0},H^{(1)}\}=\bar{V}_{\alpha_0}^{\gamma_0}\Omega_{\gamma_0}+\bar{V}_{\alpha_0}^{p_1}\Omega_{p_1}.
\end{equation}
In the other hand $\dot{\Omega}_{\alpha_0}=0$ on $\Gamma_1$. Then (\ref{21}) and expressions below, on $\Gamma_1$, means
\begin{equation}
\bar{V}_{\alpha_0}^{p_1}\Omega_{p_1}=0.
\end{equation}
Let us suppose that there are $M_0$ primary constraints $\Omega_{\alpha_0}$ and $M_1$ first-stage secondary constraints $\Omega_{p_1}$. The unique way in which the present expression implies $\Omega_{p_1}=0\,\forall\,p_1$ is that $M_1\leq M_0$ and $ \mbox{Rank} [V_{\alpha_0}^{p_1}]=M_1$ maximal. \emph{Similar assumptions on the ranks of the subsequent matrices $V_{n_s}^{n_{s+1}}$ that appear below will also be made, justified in similar arguments.}

The next step in the consistency algorithm is to demand the preservation in time of the secondary constraints $\Omega_{p_1}$ on $\Gamma_2$, being $\Gamma_2$ defined by $\Omega_{p_0}=0$ and $\Omega_{p_1}=0$. A similar reasoning leads us to
\begin{eqnarray}
\{\Omega_{p_1},\Omega_{\beta_0}^{(1)}\} & =\bar{C}_{p_1\beta_0}^{\gamma_0}\Omega_{\gamma_0}+\bar{C}_{p_1\beta_0}^{q_1}\Omega_{q_1},\nonumber\\
\{\Omega_{p_1},H^{(1)}\} & =\bar{V}_{p_1}^{\gamma_0}\Omega_{\gamma_0}+\bar{V}_{p_1}^{q_1}\Omega_{q_1}+\bar{V}_{p_1}^{p_2}\Omega_{p_2}.
\label{CM}
\end{eqnarray}
To illustrate this point we present in SubSection \ref{appD} of this appendix an example of the constructive methodology for a system that reaches the first level in the consistency algorithm.

Following the procedure, at level $(s+1)$ the consistency algorithm of the $s$-stage constraints, $\Omega_{n_s}=0$, on $\Gamma_{s+1}$ leads us to
\begin{eqnarray}
\{\Omega_{n_s},\Omega_{\beta_0}^{(1)}\} & =\sum_{i\leq s}\bar{C}_{n_s\beta_0}^{n_i}\Omega_{n_i},\\
\{\Omega_{n_s},H^{(1)}\} & =\sum_{i\leq s+1}\bar{V}_{n_s}^{n_i}\Omega_{n_i},
\end{eqnarray}
where $n_s$ denotes collectively ($\alpha_0,p_1,p_2,\ldots$). If the algorithm ends at level $L$:
\begin{equation}
\{\Omega_{n_L},H^{(1)}\}=\sum_{i\leq s}\bar{V}_{n_L}^{n_i}\Omega_{n_i}.
\end{equation}
As result of assumption (ii) and because the Poisson bracket of first-class quantities is again a first-class quantity, the equations above for first-class constraints take the form
\begin{eqnarray}
\{\Omega_{\beta_0}^{(1)},\Omega_{n_s}^{(1)}\} & =\sum_{i\leq s}C_{\beta_0 n_s}^{n_i}\Omega_{n_i}^{(1)},\nonumber\\
\{H^{(1)},\Omega_{n_s}^{(1)}\} & =\sum_{i\leq s+1}V_{n_s}^{n_i}\Omega_{n_i}^{(1)}.
\label{22}
\end{eqnarray}
If the algorithm for first-class constraints ends at level $L$:
\begin{equation}
\{H^{(1)},\Omega_{n_L}^{(1)}\}=\sum_{i\leq s}V_{n_L}^{n_i}\Omega_{n_i}^{(1)}.
\label{23}
\end{equation}
Now, we recall that the gauge transformation leaving the total action invariant implies the relations \cite{Teitelboim}
\begin{eqnarray}
\dot{\varepsilon}^{p_i}_{(1)} & =v^{\gamma_0}_{(1)}\varepsilon^{B_1}_{(1)}C_{\gamma_0 B_1}^{p_i}+\varepsilon^{B_1}_{(1)}V_{B_1}^{p_i},\label{R1}\\
\delta v^{\beta_0}_{(1)} &=\dot{\varepsilon}^{\beta_0}_{(1)}-v^{\gamma_0}_{(1)}\varepsilon^{B_1}_{(1)}C_{\gamma_0 B_1}^{\beta_0}-\varepsilon^{B_1}_{(1)}V_{B_1}^{\beta_0}.
\label{R}
\end{eqnarray}
From (\ref{22}), $C_{\beta_0 n_s}^{n_i}=0\,\forall\,i>s$ and $V_{n_s}^{n_i}=0\,\forall\,i>s+1$. Thus, we can write (\ref{R1}) in the form
\begin{equation}
\dot{\varepsilon}^{p_i}_{(1)}=v^{\gamma_0}_{(1)}\sum_{s\geq i}\varepsilon^{n_s}_{(1)}C_{\gamma_0 n_s}^{p_i}+\sum_{s\geq i-1}\varepsilon^{n_s}_{(1)}V_{n_s}^{p_i}.
\label{24}
\end{equation}
The solution of (\ref{24}) regarded as a system of equations for the $\varepsilon$'s can be constructed step by step, starting from the last expression with $i=L$ and going back until reaching $i=1$.\\
So, $i=L$ in (\ref{24}) implies
\begin{equation}
\varepsilon^{n_{L-1}}_{(1)}V_{n_{L-1}}^{p_L}=\dot{\varepsilon}^{p_L}_{(1)}-v^{\gamma_0}_{(1)}\varepsilon^{n_L}_{(1)}C_{\gamma_0 n_L}^{p_L}-\varepsilon^{n_L}_{(1)}V_{n_L}^{p_L}.
\label{25}
\end{equation}
Let us now suppose there are $m_0\equiv N_1^{(p)}$ first-class constraints at level zero, $m_1$ at level one and so on until $m_L$ at level $L$. Notice from the RHS of (\ref{25}) there are $m_L$ functions $\varepsilon^{n_L}_{(1)}$ that cannot be determined by the same equation. Assuming $\mbox{Rank}\, [V_{n_{L-1}}^{p_L}]=m_L$ maximal, Eq. (\ref{25}) determines $m_L$ of the parameters $\varepsilon^{n_{L-1}}_{(1)}$ in terms of $\varepsilon^{n_L}_{(1)}$, $\dot{\varepsilon}^{p_L}_{(1)}$, $v^{\gamma_0}_{(1)}$, $q$ and $p$. The remaining $m_{L-1}-m_{L}$ parameters are still arbitrary. Thus, at this stage, there are $m_L+(m_{L-1}-m_{L})=m_{L-1}$ arbitrary functions and $m_L$ appear together with their first time derivatives $\dot{\varepsilon}^{p_L}_{(1)}$.

At next level, $i=L-1$:
\begin{eqnarray}
\varepsilon^{n_{L-2}}_{(1)}V_{n_{L-2}}^{p_{L-1}} &=& \dot{\varepsilon}^{p_{L-1}}_{(1)} - v^{\gamma_0}_{(1)}\varepsilon^{n_{L-1}}_{(1)}C_{\gamma_0 n_{L-1}}^{p_{L-1}}-v^{\gamma_0}_{(1)}\varepsilon^{n_L}_{(1)}C_{\gamma_0 n_L}^{p_{L-1}} \nonumber \\
&& - \varepsilon^{n_{L-1}}_{(1)}V_{n_{L-1}}^{p_{L-1}}-\varepsilon^{n_L}_{(1)}V_{n_L}^{p_{L-1}}.
\end{eqnarray}
By replacing the parameters $\varepsilon^{n_{L-1}}_{(1)}$ that we solve in the previous step in $\dot{\varepsilon}^{p_{L-1}}_{(1)}$, we observe that the time derivative increase by one unit. Additionally, $\mbox{rank} \, [V^{p_{L-1}}_{p_{L-2}}]=m_{L-1}$ and then we can solve $m_{L-1}$ of the parameters $\varepsilon^{n_{L-2}}_{(1)}$ and $m_{L-2}-m_{L-1}$ remain arbitrary. In sum we have at this stage $m_L+(m_{L-1}-m_L)+(m_{L-2}-m_{L-1})$ arbitrary functions.

If we continue the procedure back until the step $i=k$, (\ref{24}) takes the form
\begin{equation}
\varepsilon^{n_{k-1}}_{(1)}V_{n_{k-1}}^{p_k}=\dot{\varepsilon}^{p_k}_{(1)}-v^{\gamma_0}_{(1)} 
\sum_{j\geq k}\varepsilon^{n_j}_{(1)}C_{\gamma_0 n_j}^{p_k}-\sum_{j\geq k}\varepsilon^{n_j}_{(1)}V_{n_j}^{p_k}.
\label{26}
\end{equation}
Replacing the parameters already solved in steps $i>k$, we reach the following situation:
The RHS of (\ref{26}) depends on $q$'s, $p$'s, $v^{\alpha_0}_{(1)}$ and their time derivatives up to order $L-k+1$ introduced by the term $\dot{\varepsilon}^{n_{i}}_{(1)}$ step by step, as well as on the remaining arbitrary function $\varepsilon^{n_{s}}_{(1)}$ $s\geq k$ introduced in the analysis for $i>k$, and their time derivatives.

Because $V_{n_{k-1}}^{p_k}$ is of maximal rank $m_k$, we determine $m_k$ of the $m_{k-1}$ functions $\varepsilon^{n_{k-1}}_{(1)}$ and the remaining $m_{k-1}-m_k$ still arbitrary. Thus, at this stage we increase the order of time derivatives of the previously introduced gauge parameters by one unit and introduce $m_{k-1}-m_k$ new arbitrary functions.\\
If we continue the procedure until reach the stage $i=1$, we end with a number of totally arbitrary functions of
\begin{eqnarray}
&& m_L+(m_{L-1}-m_L) + \ldots +(m_{k-1}-m_k) + \nonumber\\
&& + \ldots +(m_0-m_1)=m_0\equiv N_1^{(p)}.
\label{27}
\end{eqnarray}
Since the Lagrangian and the Hamiltonian gauge transformation law $\delta_\varepsilon F\approx\{F,\varepsilon^A_{(1)}\Omega_A^{(1)}\}$ have to be equivalents,\cite{Zanelli} they must have the same quantity of arbitrary independent functions of time. Thus,
\begin{equation}
N_1^{(p)}=g.
\end{equation}
This proves the Result 1.

On the other hand, the first $m_L$ arbitrary parameters appear with derivatives up to order $L+1$, the following $m_{L-1}-m_L$ up to $L$ and so on until the last $m_0-m_1$ gauge parameters appear undifferentiated. The total number of parameters plus their successive time derivatives is
\begin{eqnarray}
e & = & (L+1)m_L+L(m_{L-1}-m_L)+\ldots +k(m_{k-1}-m_k)+ \nonumber \\
& &(k-1)(m_{k-2}-m_{k-1})+\ldots +2(m_1-m_2)+(m_0-m_1) \nonumber \\
&=& m_L+m_{L-1}+\ldots +m_1+m_0 \nonumber \\
&\equiv & N_1,
\label{Dan1}
\end{eqnarray}
which completes the proof of the Result 2.
\subsection{\label{appD}Illustrating the constructive methodology, an example}
In order to get some insight about the sequential pattern described in (\ref{CM}) we want to illustrate it with an example. We will take the example considered in the SubSection \ref{sec4.1}. The Lagrangian action is characterized by the action (\ref{30})
\begin{equation*}
L(q^i, \dot{q}^i)=\dot{q}^1 q^2-\dot{q}^2 q^1- (q^1-q^2) q^3.
\end{equation*}
The Hamiltonian analysis reported in Ref.~\onlinecite{Chitaia} involves three \emph{primary} constraints $(\alpha_0=\{1,2,3\})$
\begin{equation}
\Omega_1:= p_1- q^2\approx 0, \quad
\Omega_2 := p_2+ q^1\approx 0, \quad 
\Omega_3 := p_3\approx 0,
\end{equation}
and a \emph{primary} Hamiltonian $H_0=(q^1-q^2)q^3$. The evolution of the primary constraints using the \emph{total Hamiltonian} $H_T=H_0+v^{\alpha_0}\Omega_{\alpha_0}$  gives the \emph{secondary} constraint $(p_1=4)$
\begin{equation}
\Omega_4:= q^2-q^1\approx 0.
\end{equation}
Classifying the constraints, the \emph{first-class} ones are
\begin{equation*}
\Omega_1^{(1)}:=\Omega_3=p_3, \quad
\Omega_2^{(1)}:=\Omega_1+\Omega_2+2, \quad \Omega_4=p_1+p_2+q^2-q^1, \quad
\end{equation*}
and the \emph{second-class} constraints can be chosen as
\begin{equation}
\Omega_1^{(2)}:=\Omega_1, \quad 
\Omega_2^{(2)}:=\Omega_2.
\end{equation}
Thus, $N_1=2$ and $N_2=2$. If we compare with the Lagrangian analysis of the current approach carried out in SubSection \ref{sec4.1}. is easy to see that $N_1+N_2=4\equiv l+g$, which is consistent with our general result.

Now, computing the \emph{first-class Hamiltonian}:
\begin{equation}
H^{(1)}=\frac{q^3}{2}(q^1-q^2-p_1-p_2),
\end{equation}
and then $H_T=H^{(1)}+v^{1}_{(1)}\Omega^{(1)}_1$.

We already have all the constraints and we have classified them. Now, we will see how the Poisson brackets that we need for the evolution look like. First notice that
\begin{equation*}
\dot{\Omega}_{\alpha_0}=\{\Omega_{\alpha_0} ,H^{(1)}\}+v^{1}_{(1)}\{\Omega_{\alpha_0},\Omega_{1}^{(1)}\}\approx 0,
\end{equation*}
and then we need $\{\Omega_{\alpha_0},\Omega_{1}^{(1)}\}$ and $\{\Omega_{\alpha_0},H^{(1)}\}$. Computing
\begin{equation}
\left\{\Omega_1,\Omega_1^{(1)}\right\}=\left\{\Omega_2,\Omega_1^{(1)}\right\}=\left\{\Omega_3,\Omega_1^{(1)}\right\}=0\,\,\Longrightarrow \bar{C}_{\alpha_0\beta_0}^{\gamma_0}=0,
\end{equation}
and
\begin{equation}
\left.
\begin{array}{l}
\left\{H^{(1)},\Omega_1\right\}=\left\{H^{(1)},\Omega_2\right\}=0,\\
\left\{H^{(1)},\Omega_3\right\}=-\frac{1}{2}\Omega_1-\frac{1}{2}\Omega_2-\Omega_4,
\end{array}
\right\}\Longrightarrow
\begin{array}{l}
\bar{V}_3^1=-\frac{1}{2}=\bar{V}_3^2,\\
\bar{V}_3^{p_1} =-1,
\end{array}
\label{A1}
\end{equation}
where $p_1=4$ and the other coefficients vanish. As we see in the second line of (\ref{A1}), the Poisson bracket $\left\{H^{(1)},\Omega_3\right\}$ is a combination of the \emph{primary} constraints and the \emph{secondary} constraint generated by the method at the actual (first) order in the consistency procedure. Because $\Omega_3= \Omega_1^{(1)}$ is a \emph{first-class} constraint, $\left\{H^{(1)},\Omega_3\right\}$ is a strong combination of first-class constraints. In fact, $\left\{H^{(1)},\Omega_3\right\}=-\frac{1}{2}\Omega_2^{(1)}$. This is of the same type as the general case expressed by Eq. (\ref{22}).

At next order in the consistency method
\begin{equation*}
\dot{\Omega}_{p_1}=\{\Omega_{p_1} ,H^{(1)}\}+v^{1}_{(1)}\{\Omega_{p_1},\Omega_{1}^{(1)}\}\approx 0,
\end{equation*}
and then we need to compute $\{\Omega_{p_1} ,H^{(1)}\}$ and $\{\Omega_{p_1},\Omega_{1}^{(1)}\}$. In this case
\begin{equation}
\begin{array}{l l}
\{\Omega_4 ,\Omega_1^{(1)}\}=0&\Longrightarrow \bar{C}_{p_1\beta_0}^{\gamma_0}=0=\bar{C}_{p_1\beta_0}^{q_1},\\
\{\Omega_4 ,H^{(1)}\}=0&\Longrightarrow \bar{V}_{p_1}^{\gamma_0}=\bar{V}_{p_1}^{q_1}=\bar{V}_{p_1}^{p_2}=0,
\end{array}
\end{equation}
which implies that the algorithm ends at level $L=1$.
\subsubsection{Reduction procedure}
As we saw above, the evolution of the primary first-class constraint $\Omega_1^{(1)}$ requires to compute
\begin{equation}
\begin{array}{l l}
\left\{\Omega_1^{(1)},\Omega_1^{(1)}\right\}=0 & \Longrightarrow C_{\alpha_0\beta_0}^{\gamma_0}=0,\\
\left\{\Omega_1^{(1)}, H^{(1)}\right\}=-\frac{1}{2}\Omega_2^{(1)} & \Longrightarrow V_{\alpha_0}^{\gamma_0}=0,\,\,V_1^2=-\frac{1}{2}.
\end{array}
\label{A2}
\end{equation}
The evolution of the new constraint $\Omega_2^{(1)}$ requires
\begin{equation}
\begin{array}{l l}
\left\{\Omega_2^{(1)},\Omega_1^{(1)}\right\}=0 & \Longrightarrow C_{p_1\beta_0}^{\gamma_0}=0=C_{p_1\beta_0}^{q_1},\\
\left\{\Omega_2^{(1)}, H^{(1)}\right\}=0 & \Longrightarrow V_{p_1}^{\gamma_0}=V_{p_1}^{q_1}=V_{p_1}^{p_2}=0.
\end{array}
\label{A3}
\end{equation}
Thus the algorithm stops here, at level $L=1$, as expected.

Applying the Eq. (\ref{25}) for $L=1$ we get
\begin{eqnarray}
\varepsilon^{n_0}_{(1)}V_{n_0}^{p_1}=\dot{\varepsilon}^{p_1}_{(1)}-v^{\gamma_0}_{(1)}\varepsilon^{n_1}_{(1)}C_{\gamma_0 n_1}^{p_1}-\varepsilon^{n_1}_{(1)}V_{n_1}^{p_1}.
\label{A4}
\end{eqnarray}
With the information in (\ref{A2})-(\ref{A3}), the Eq. (\ref{A4}) takes the form
\begin{eqnarray}
\varepsilon^1_{(1)}V_1^2=\dot{\varepsilon}^2_{(1)}\Longrightarrow \varepsilon^1_{(1)}=-2\dot{\varepsilon}^2_{(1)},
\end{eqnarray}
where we realize that we reduce by one the number of free parameters and increase by one the order of the time derivative, in agreement with the general discussion of this appendix. If we had had more levels $L$ in the consistency procedure we would have to continue from the later $i=L$ until get $i=1$ in a closed form to the one described here. Let us pointed out that in the actual case $e=2=N_1$ and $g=1=N_1^{(p)}$, which is a general result.

\section{\label{appC}A counterexample to Dirac's conjecture}
One of the main statements used in the proof of the formula for the physical degree of freedom count (\ref{29}) was that Lagrangian and Hamiltonian gauge transformations should be equivalents in the total formalism. However it is well known this is not true in counterexamples to Dirac's conjecture and it is natural to ask if there are any changes to (\ref{29})? Let us study a particular example defined by the Lagrangian
\begin{equation}
L=\frac{1}{2}e^y\dot{x}^2,
\label{C1}
\end{equation}
which is a counterexample to Dirac's conjecture.\cite{Teitelboim}
\subsection{Hamiltonian analysis}\label{B.1}
For the sake of completeness, and in order to compare with the Lagrangian approach carried out in Subsection \ref{B.2} of this Appendix, here we review the Dirac's Hamiltonian formalism of the Lagrangian (\ref{C1}), which is reported in Ref. \onlinecite{Teitelboim}.
The momenta canonically conjugate to $(x,y)$ are given by
\begin{equation}
p_x=\dot{x}e^y,
\hspace{1cm}
p_y=0,
\label{C2}
\end{equation}
which imply the primary constraint $\gamma_1 := p_y\approx 0$ and
the primary Hamiltonian is
\begin{equation}
H_0=\frac{1}{2}p_x^2e^{-y}.
\label{C3}
\end{equation}
The evolution of the primary constraint leads to the secondary constraint satisfying the regularity condition $\gamma_2:= p_x\approx 0$. The consistency algorithm applied to $\gamma_2$ does not lead to new Hamiltonian constraints and the procedure stops here. It is worth noting that even though the election of constraints $\Gamma_1=\gamma_1$ and $\Gamma_2=(\gamma_2)^2$ made in Ref.~\onlinecite{Kiriushcheva} leads to the same (Hamiltonian) gauge transformation as that coming from the Lagrangian
analysis, that choice of the constraints {\it does not} satisfy the regularity condition.\cite{Teitelboim}

It is clear that $\{\gamma_1, \gamma_2 \} =0$ and therefore all constraints are \emph{first-class}: $N_1=2$, $N_2=0$ and the number of physical degrees of freedom is
\begin{equation}
2-2-\frac{0}{2}=0.
\label{C6}
\end{equation}
Following Dirac's conjecture, all constraints are generators of gauge transformations, which are
\begin{eqnarray}
&&\delta_{\varepsilon} x = {\varepsilon}^2, \quad \delta_{\varepsilon} y = {\varepsilon}^1, \nonumber\\
&& \delta_{\varepsilon} p_x = 0, \quad \delta_{\varepsilon} p_y =0,
\label{C4}
\end{eqnarray}
where $\varepsilon^1$ and $\varepsilon^2$ are independent arbitrary gauge parameters. The total Hamiltonian formalism can be obtained by setting to zero the Lagrangian multipliers related to first-class secondary constraints in the extended action, and this has as consequence relations (\ref{R1}) and (\ref{R}). Since $\left\{H_0,\gamma_1\right\}=-\frac{1}{2}\dot{x}\gamma_2\approx 0$ and $\left\{H_0,\gamma_2\right\}=0$ we have that $H_0$ is a first-class quantity, $V_1^2=-\frac{1}{2}\dot{x}$ and the others components of the matrix $V$ vanish. Additionally, $\left\{\gamma_1,\gamma_2\right\}=0$ implies that $C$ is the zero matrix. Inserting in (\ref{R1}): $\dot{\varepsilon}^2=\varepsilon^1V_1^2\equiv -\frac{1}{2}\dot{x}\varepsilon^1$ and therefore $\varepsilon^1=-2\dot{\varepsilon}^2/\dot{x}$. Defining $\varepsilon := \varepsilon^2$ we get in the total formalism
\begin{equation}
\delta_{\varepsilon}^{total} x=\varepsilon,
\hspace{0.5cm}
\delta_{\varepsilon}^{total} y=- \frac{2 {\dot \varepsilon}} {\dot x},
\hspace{0.5cm}
\delta_{\varepsilon}^{total} p_x=\delta_{\varepsilon}^{total} p_y=0.
\label{C5}
\end{equation}
\subsection{Analysis using the approach of the Section \ref{sec3}}\label{B.2}
The variational derivatives
\begin{equation}
\begin{array}{l l}
E^{0}=\left(
\begin{array}{c}
E_1^{0}\\
E_2^{0}
\end{array}
\right) & =\left(
\begin{array}{c}
(\ddot{x}+\dot{x}\dot{y})e^y\\
-\frac{1}{2}\dot{x}^2e^y
\end{array}
\right)\\
         & =W^0\left(
        \begin{array}{c}
        \ddot{x}\\
        \ddot{y}
        \end{array}
        \right)+K^0,
\end{array}
\label{C8}
\end{equation}
with
\begin{equation}
W^{0}=\left(
\begin{array}{c c}
e^y & 0\\
0 & 0
\end{array}
\right),
\hspace{1cm}
K^{0}=\left(
\begin{array}{c}
\dot{x}\dot{y}e^y\\
-\frac{1}{2}\dot{x}^2e^y
\end{array}
\right).
\label{C7}
\end{equation}
The solution to the Euler-Lagrange equations, $E^0=0$, is $x=x_0= \mbox{cte}$ and $y$ arbitrary. On the other hand, (\ref{C5}) means that there are arbitrary shifts in $x$ that do not correspond with any arbitrariness in the general solution $x=x_0$ and therefore $\gamma_2$ is generating gauge transformations that do not correspond with the Lagrangian dynamics (counterexample).\cite{Teitelboim}

A basis for the left null vectors of $W^0$ is
\begin{eqnarray}
\lambda_1 = (0,1), \label{plus}
\end{eqnarray}
and contracting (\ref{C8}) with (\ref{plus}) gives
\begin{eqnarray}
E^0_2 = K^0_2.
\label{C9}
\end{eqnarray}
This is not a gauge identity. From (\ref{C9}) and $E^0=0$, we get the constraint
\begin{eqnarray}\label{masmas}
\psi_0 := K^0_2 =0.
\end{eqnarray}
End of step 0, which includes $g_0 =0$ gauge identities and $l_0=1$ Lagrangian constraints.

Step 1. We add ${\dot \psi}_0=0$ to $E^0=0$
\begin{equation}
\begin{array}{l l}
E^{1}=\left(
\begin{array}{c}
E^{0}\\
\frac{d}{dt} \psi_0
\end{array}
\right) & =\left(
\begin{array}{c}
(\ddot{x}+\dot{x}\dot{y})e^y\\
-\frac{1}{2}\dot{x}^2e^y\\
-(\dot{x}\ddot{x}+\frac{1}{2}\dot{x}^2\dot{y})e^y
\end{array}
\right)\\
         & =W^1\left(
         \begin{array}{c}
         \ddot{x}\\
         \ddot{y}
         \end{array}
         \right)+K^1,
\end{array}
\label{C10}
\end{equation}
where
\begin{equation}
W^{1}=\left(
\begin{array}{c c}
e^y & 0\\
0 & 0\\
-\dot{x}e^y & 0
\end{array}
\right),
\hspace{1cm}
K^{1}=\left(
\begin{array}{c}
\dot{x}\dot{y}e^y\\
-\frac{1}{2}\dot{x}^2e^y\\
-\frac{1}{2}\dot{x}^2\dot{y}e^y
\end{array}
\right).
\label{C11}
\end{equation}
A basis for the left null vectors of $W^1$
\begin{eqnarray}
\lambda_1 = \left ( 0,1,0 \right ), \quad \lambda_2 = \left ({\dot x}, 0,1  \right ). \label{plus2}
\end{eqnarray}
$\lambda_1$ is (\ref{plus}) with an additional zero entry and by contracting $E^1$  with it we get (\ref{C9}). On the other hand, contracting
$E^1$ with $\lambda_2$
\begin{eqnarray}
\dot{x}E^{0}_1+\frac{d}{dt} \psi_0 &=\frac12 {\dot x}^2 {\dot y} e^y
\label{C12}
\end{eqnarray}
Using (\ref{C9}) and (\ref{masmas}) we rewrite $\psi_0$ as $\psi_0=E^0_2$ and identifying  in the RHS of (\ref{C12}) $-\frac12 {\dot x}^2 e^y=E^0_2$, we rewrite (\ref{C12}) as
\begin{equation}
G^1:= \dot{x}E^{0}_1+\frac{d}{dt}E^{0}_2+\dot{y}E^{0}_2=0.
\label{C13}
\end{equation}
This is a \emph{gauge identity}. The procedure ends here at step 1, which includes $g_1=1$ gauge identities and $l_1=0$ Lagrangian constraints.

Contracting (\ref{C13}) with an arbitrary parameter $\varepsilon(t)$ and rewriting we get the Noether's identity
\begin{equation}
\dot{x}\varepsilon E^{0}_1+(\dot{y}\varepsilon-\dot{\varepsilon})E^{0}_2+\frac{d}{dt}\left ( \varepsilon E^{0}_2 \right ) =0,
\label{C14}
\end{equation}
where we can directly read the gauge transformation law for the coordinates
\begin{equation}
\delta_\varepsilon x=\dot{x}\varepsilon\,\,,
\hspace{1cm}
\delta_\varepsilon y=\dot{y}\varepsilon-\dot{\varepsilon}.
\label{C15}
\end{equation}
In summary, we have 2 original variables ($x$ and $y$), $l=l_0 + l_1 = 1 +0 =1$ independent Lagrangian constraints, $g=g_0 + g_1 = 0+1 =1$ gauge identity and $e=2$ effective gauge parameters (${\varepsilon}$ and ${\dot \varepsilon}$). Using the expression (\ref{29}), the number of physical degrees of freedom is
\begin{equation}
2-\frac{1}{2}(1+1+2)=0,
\label{C16}
\end{equation}
which is the same that we get in the Hamiltonian analysis. Furthermore, relations (\ref{key2}) and (\ref{key3}) give $N_1=2$ first-class constraints and $N_2=0$ second-class constraints, in agreement with the Hamiltonian analysis.

It is worth mentioning that the gauge identity (\ref{C13}) agrees with the one found in Ref. \onlinecite{Kiriushcheva}. Nevertheless, in  Ref. \onlinecite{Kiriushcheva} it was obtained from the knowledge of the gauge transformation of the Hamiltonian analysis. Here, in opposition, we generate this gauge identity from the Lagrangian formalism only and then we find the gauge transformation (\ref{C15}), avoiding the Hamiltonian analysis.

From (\ref{C5}) and (\ref{C15}) it is not clear that \emph{total} and \emph{Lagrangian} gauge transformation laws are equivalent. In fact, they are not equivalent since requiring $\delta_{\varepsilon}^{total} x=\delta_\varepsilon x$ and $\delta_{\varepsilon}^{total} y=\delta_\varepsilon y$ we run into difficulties ($\dot{x}$ is no longer zero on shell). This is consistent with the claim made in Ref.~\onlinecite{Wang}, even when the calculations are wrong there.
As a matter of fact this example shows that, even getting different gauge transformation laws in both treatments, the direct application of the original theory gives the right physical information and reinforce the fundamental character of the construction.

\subsection{The approach from the covariant canonical formalism}\label{B.3}
The Lagrangian (\ref{C1}) leads to the presymplectic two-form
 $\Omega$ and the energy $E$
\begin{eqnarray}
\Omega &= &  e^y \dot{x} dx \wedge dy + e^y dx \wedge d \dot{x}, \nonumber \\
 E&=& \frac{1}{2} e^y \dot{x}^2, \nonumber  \\
\Rightarrow dE&=& e^y \dot{x} \left( \frac{\dot{x}dy}{2} + d\dot{x}  \right).
\end{eqnarray}
A basis of $\ker \Omega$ is
$\displaystyle \left\{ \frac{\partial}{\partial y}-\dot{x}\frac{\partial}{\partial \dot{x}},
\frac{\partial}{\partial \dot{y} }  \right\}$. Notice that only the first one generates a Lagrangian constraint, given by $\phi= - \frac{\dot{x}^2}{2} e^y=0$. We now demand that $X (\phi)=0$,
but because $X$ satisfies (\ref{gp2}) this is accomplished.
Therefore, the constraint algorithm gives us just one Lagrangian constraint which is the projectable one.

Following the procedure, Eqs. (\ref{gp2}) become
\begin{eqnarray}
-(\dot{x}\alpha^2+ \beta^1)e^y& = & 0,  \nonumber\\
e^y \dot{x} (\alpha^1- \frac{\dot{x}}{2})   & = & 0, \nonumber \\
e^y(\alpha^1- \dot{x}) &=&0, \label{C18}
\end{eqnarray}
and the requirement $\alpha^1=\dot{x}$, $\alpha^2=\dot{y}$ implies that $\dot{x}$ must satisfy
\begin{eqnarray}
\dot{x}=0,
\end{eqnarray}
which is consequence of $\phi$. Thus, there are not any non-projectable constraints.

Therefore, we have $l=1$ Lagrangian constraints (in agreement with the result of SubSection \ref{B.2}) and if we restrict $\Omega$ to them, we have
\begin{eqnarray}
\Omega|_{\phi}=0,
\end{eqnarray}
and the number of physical degrees of freedom is
\begin{eqnarray}
\frac{1}{2} \mbox{Rank}\,\, \Omega|_{\phi}=0.
\end{eqnarray}
In order to get the gauge transformation, we
compute
\begin{eqnarray}
0=(\tilde{X}^j \contr \,\, \Omega)|_{\phi}= e^y \delta_\varepsilon \dot{x}dx.
\end{eqnarray}
Therefore $\delta_\varepsilon \dot{x}=0$, and we do not get information on $\delta_\varepsilon y$. This transformation law is, of course, in agreement with (\ref{C15}) {\it on-shell}. Why do not we get (\ref{C15})? It is because we get a degenerate direction over the space of solutions.

\nocite{*}

\end{document}